\documentclass [prb, preprint,notitlepage,superscriptaddress,floatfix] {revtex4-1}
\usepackage{amsmath}
\usepackage{graphicx}
\usepackage{lmodern}
\usepackage{amsmath}
\usepackage{xcolor}

\definecolor{darkgreen}{RGB}{0,170,0}

\usepackage{amssymb}
\usepackage{bm}
\usepackage{ulem}
\newcommand{\beq} {\begin{equation}}
\newcommand{\eeq} {\end{equation}}
\newcommand{\bea} {\begin{eqnarray}}
\newcommand{\eea} {\end{eqnarray}}
\newcommand{\be} {\begin{equation}}
\newcommand{\ee} {\end{equation}}
\renewcommand{\(}{\left(}
\renewcommand{\)}{\right)}

\begin{document}

\title {Interplay between superconductivity and non-Fermi liquid at a quantum critical point in a metal.
II.  The $\gamma$-model at a finite $T$ for $0<\gamma <1$.}
\author{Yi-Ming Wu}
\affiliation{School of Physics and Astronomy and William I. Fine Theoretical Physics Institute, University of Minnesota, Minneapolis, MN 55455, USA}
\author{Artem Abanov}
\affiliation{Department of Physics, Texas A\&M University, College Station,  USA}
\author{Yuxuan Wang}
\affiliation{Department of Physics, University of Florida, Gainesville, FL 32611,  USA}
\author{Andrey V. Chubukov}
\affiliation{School of Physics and Astronomy and William I. Fine Theoretical Physics Institute,
University of Minnesota, Minneapolis, MN 55455, USA}
\date{\today}

\begin{abstract}
In this paper we continue the analysis of the interplay
 between non-Fermi liquid and superconductivity for
  quantum-critical systems, the low-energy physics of which is described by
  an effective model with dynamical electron-electron interaction
 $V(\Omega_m) \propto 1/|\Omega_m|^\gamma$ (the $\gamma$ model). In paper I [A. Abanov and A. V. Chubukov, arXiv:2004.13220 (2020)]
    two of us analyzed the $\gamma$ model at $T=0$ for $0<\gamma <1$ and argued that there exist a discrete, infinite set of topologically distinct solutions for the superconducting gap, all with the same spatial symmetry. The gap function  $\Delta_n (\omega_m)$ for the $n$th solution changes sign $n$ times as the function of Matsubara frequency.
    In this paper we analyze the linearized gap equation at a finite $T$. We show that there exist an infinite set of pairing instability temperatures, $T_{p,n}$, and the eigenfunction $\Delta_n (\omega_{m})$
       changes sign $n$ times as a  function of a Matsubara number $m$. We argue that $\Delta_n (\omega_{m})$
    retains its functional form below $T_{p,n}$ and at $T=0$
     coincides with the $n$th solution of the nonlinear gap equation.
      Like in paper I, we extend the model to the case when the interaction in the pairing channel has  an additional factor $1/N$ compared to that in the particle-hole channel.
     We show that $T_{p,0}$  remains finite at large $N$
      due to special properties of fermions with Matsubara frequencies $\pm \pi T$,
      but all other $T_{p,n}$
         terminate
         at $N_{cr} = O(1)$.   The gap function vanishes at $T \to 0$ for $N > N_{cr}$  and remains finite for $N < N_{cr}$. This is consistent with the $T =0$ analysis.
\end{abstract}
\maketitle

\newpage

\section{Introduction}

In this paper we continue our analysis of the competition between non-Fermi-liquid (NFL) physics and superconductivity near a quantum-critical point (QCP) in a metal for a class of models with dynamical four-fermion interaction $V(q, \Omega_m)$,  mediated by a critical soft boson.  The interaction $V(q, \Omega_m)$ gives rise to strong fermionic self-energy, which destroys Fermi-liquid behavior, in most cases in dimensions $D \leq 3$, and also gives rise to an attraction in at least one pairing channel.  We consider the  class of models in which bosons are slow excitations compared to fermions (like phonons in the case when the Debye frequency is much smaller than the Fermi energy).  In this situation, the momentum integration in the expressions for the fermionic self-energy and the pairing vertex in the proper attractive channel can be carried out, and both the self-energy  and the pairing vertex  are determined by the effective, purely dynamical local interaction $V (\Omega_m) \propto \int d{\bf q} V(q, \Omega)$, with ${\bf q}$ connecting points on the Fermi surface.  The coupled equations for
the self-energy $\Sigma (k_F, \omega_m)$ and the pairing vertex $\Phi (\omega_m)$ have the same structure as Eliashberg equations for a phonon superconductor.  We consider  quantum-critical models for which $
V (\Omega_m) = {\bar g}^\gamma/|\Omega_m|^\gamma$ (the $\gamma$ model). In the previous paper, Ref. \cite{paper_1}, hereafter  referred to as paper I, we listed a number of quantum-critical systems, the low-energy physics of which is described by the $\gamma$ model with different $\gamma$.  This paper also contains an extensive list of references to earlier publications on this subject.

The interaction $V(\Omega_m)$ is singular, and gives rise to two
 opposite
 tendencies: NFL
  behavior  in the normal state, with
$\Sigma (\omega_m) \propto \omega^{1-\gamma}_m$, and the pairing. The two tendencies compete with each other as
  a NFL self-energy reduces the magnitude of the pairing kernel, while the feedback from the pairing reduces fermionic self-energy.

In paper I we analyzed zero-temperature behavior of the $\gamma$ model for $0 <\gamma <1$.  We found that the
 system does become unstable towards pairing, i.e., in the ground state
 the pairing vertex $\Phi (\omega_m)$ and the pairing gap $\Delta (\omega_m) = \Phi (\omega_m)/(1+ \Sigma (\omega_m)/\omega_m)$ are finite. However, in qualitative distinction with BCS/Eliashberg theory of superconductivity, in which there is a single
  solution of the gap equation, here we found an infinite discrete set of solutions $\Delta_n (\omega_m)$. The solutions have the same spatial gap symmetry, but
  %are topologically distinct as
  $\Delta_n (\omega_m)$ changes sign $n$ times as a function of Matsubara frequency. The gap functions $\Delta_n (\omega_m)$ tend to finite values at zero frequency, but the magnitude of $\Delta_n (0)$ decreases with $n$ and at large enough $n$ scales as
   $\Delta_n (0) \propto e^{-A n}$, where $A$ is a function of $\gamma$.
    The solutions with different $n$ are topologically different. To see this, one has to consider positive
     $\omega_m$ and analytically continue
the function $\Delta_{n}(\omega_{m})$
into the upper half-plane of frequency ($\omega_m \to -iz$). The gap function $\Delta_n (z)$ is a complex function for a generic $z$, $\Delta(z) = \Delta' (z) + i \Delta^{''} (z)$, and one can define $z-$dependent phase as $\Delta (z) = |\Delta (z)| e^{i \psi (z)}$.  One can verify that $\psi (z)$ winds up by $2\pi$ upon circling every nodal point on the Matsubara axis, i.e., $\Delta_n (z)$ is a solution with $n$ vortices. This same physics can be detected along the real axis ($z = \omega$):
      \footnote{This issue will be discussed in more detail  for the case $\gamma >1$, where  additional vortices appear in the complex plane for all $\Delta_n(z)$.} the total change of the phase between $\omega = - \infty$ and $\infty$ for $\Delta_n (\omega)$ has extra $2\pi n$ compared to the case $n=0$.
        \footnote{Topological aspects of the existence of zeros of the gap function in momentum space have been extensively discussed in the literature (see e.g., Refs. \cite{Sato_2017,Zhang_2019,Matsuura_2013}).
         Our case is a dynamical version of a nodal topological superconductor. The classification and consequences of dynamical nodal topology in frequency space remain an open question. }

 In this paper we analyze the  $\gamma$ model for $0 <\gamma <1$ at  finite $T$. We solve the linearized gap equation and show that there is an infinite, discrete set of the onset temperatures for the pairing
 $T_{p,n}$.  The magnitude of  $T_{p,n}$ decreases with $n$ and at large $n$,  $T_{p,n} \propto e^{-A n}$ with the same $A$ as for $\Delta_n (0)$ from the  $T=0$ analysis.   The gap function $\Delta_n (\omega_m)$ at $T= T_{p,n} -0$ changes sign $n$ times as a function of discrete Matsubara frequency $\omega_m = \pi T (2m+1)$.  We argue that $\Delta_n (\omega_m)$, which develops at  $T_{p,n}$, retains its functional form with $n$ sign changes also at $T < T_{p,n}$, and  at $T \to 0$ approaches the $n$ th solution of the full nonlinear gap equation, which we obtained in paper I.

 In paper I we also added an extra parameter $N$ to  the $\gamma$ model to control the ratio of
  interactions in the particle-particle and the particle-hole channels. Specifically, we added the factor $1/N$ to the interaction in the pairing channel and left the interaction in the particle-hole channel intact.  We found that at $T=0$ there exists a critical $N_{cr} >1$, which separates the NFL  ground state for $N > N_{cr}$ and
   the state with a nonzero $\Delta$  for $N < N_{cr}$.  We found that all $\Delta_n (\omega_m)$ emerge simultaneously at $N  = N_{cr}-0$. Here  we analyze the extended $\gamma$ model at a finite $T$.
   We show that the onset temperature $T_{p,0}$ for sign-preserving gap function $\Delta_0 (\omega_m)$ remains finite for all $N$ and
    at large $N$ scales as $T_{p,0} \propto 1/N^{1/\gamma}$.  This $T_{p,0}$ has been studied before~\cite{Wang2016}, and its existence even for very large $N$
     was  attributed to special properties of fermions with $\omega_m = \pm \pi T$, for which nonthermal part of the self-energy vanishes.
    Namely, it was argued that  the pairing at $T_{p,0}$ at large $N$ is predominantly between these fermions,  and $\Delta_{n=0} (\pm \pi T)$ is much larger than at other $\omega_m$.

   Here, we show that there exists an infinite set of other $T_{p,n}$ with $n >0$, which  terminate at
    $N = N_{cr}$ and are related to the solutions of the nonlinear gap equation at $T=0$, $\Delta_{n>0} (\omega_m)$. We show that the structure of  the
 eigenfunctions
$\Delta_{n>0} (\omega_m)$ at $T_{p,n}$
      is opposite to that of $\Delta_0 (\omega_m)$ in the sense that $\Delta_{n>0} (\omega_m)$
     has additional smallness at $\omega_m = \pm \pi T$.  We show that for any $N < N_{cr}$, $T_{p,n}$ gradually decrease with $n$, and the set become more dense as $n$ increases.  As the consequence, in the limit  $T \to 0$, the  density of  eigenvalues  (DoE) remains finite for all $N < N_{cr}$.   We  show that it is the same at $T=0$ and at $T \to 0$, i.e., the system evolves smoothly between $T=0$ and $T >0$.    For completeness, we obtain the DoE away from a QCP, when a pairing boson  has a finite mass. We show that in this case the solutions at $T \to 0$ exist only for a discrete set of $N$.

   The structure of the paper is the following.  In Sec. \ref{sec:model} we present Eliashberg equations for the $\gamma$ model and discuss in some detail the  extension to $N \neq 1$ at a finite $T$. In Sec. \ref{sec:analytics} we analyze analytically and numerically the linearized gap equation at a finite $T$. We show that there are two possibilities to get a solution at small $T$: one can either set $N \propto 1/T^\gamma$ (the $n =0$ solution), or keep $N$ finite.  We consider these two cases separately in Secs. \ref{sec:Nlarge} and \ref{sec:N_1}.  In Sec. \ref{sec:Nlarge} we find the analytic solution for $T_{p,0}$, which reproduces earlier results in Refs.\cite{Wang2016,Wu_19_1,*Abanov_19,*Chubukov_2020a}.  In Sec. \ref{sec:N_1}
    we first solve numerically the gap equation at $T \to 0$, obtain the density of eigenvalues, and show that it is nonzero for all $N$ between $N=N_{cr}$ and $0$.  This is consistent with the $T=0$ result in paper I.
  We then extend calculations to finite $T$ and show that for any $N < N_{cr}$ (and any $\gamma$ from $0<\gamma <1$), there exists an infinite, discrete  set of onset temperatures for the pairing, $T_{p,n}$.
     We argue that all $T_{p,n}$ with $n >0$ terminate at $N = N_{cr}$ and that $T_{p,n}$ at large $n$ should decay exponentially with $n$ as  $T_{p,n} \propto e^{-A n}$. We
   relate $A$ to the form of $\Delta_n (\omega_m)$ at $T=0$, express $A$ via $\gamma$ and $N$,  and argue that
    $T_{p,n}$ is proportional to $\Delta_n (0)$ at $T=0$.
   In Sec. \ref{sec:gap} we present numerical results for the gap function $\Delta_n (\omega_m)$ at $T = T_{p,n} -0$. We show that $\Delta_n (\omega_m)$ changes sign $n$ times as a function of discrete Matsubara frequency. We extend the numerical analysis to a range of $T \leq T_{p,n}$, where $\Delta_n (\omega_m)$ is small but finite, and show that $\Delta_n (\omega_m)$ still changes sign $n$ times, i.e., its topological structure does not change below $T_{p,n}$.
    \footnote{Although $\Delta_n (\omega_m)$ at $T>0$ does not necessary has nodes at the set of Matsubara points, the topological difference between the solutions with different $n$ still exists because, e.g., on a real axis the total change of the phase  of a complex $\Delta_n (\omega) = |\Delta_n (\omega)|e^{i\psi_n (\omega)}$ between $\omega = -\infty$ and $\omega = \infty$ has extra $2\pi n$ compared to the case $n=0$.}

    We argue that at $T \to 0$ this $\Delta_n (\omega_m)$ approaches the $n$th solution of the nonlinear gap equation at $T=0$, which has the same topological structure.
   In Sec. \ref{sec:away_from_a_qcp} we discuss how the density of eigenvalues at $T \to 0$ and the onset temperatures $T_{p,n}$ evolve away from a QCP, when the pairing boson acquires a finite mass $M$.
   Finally, in  Sec. \ref{sec:conclusions} we present our conclusions and briefly outline what we will do next.

Before we proceed, we briefly outline our reasoning to look for solutions with different $\Delta_n$.
For $\gamma <1$, which we study here, the set of temperatures $T_{p,n}$
is infinite, but discrete, and $T_{p,0}$ for $n=0$ is the largest onset temperature for the pairing. there is just one phase transition, at the largest $T_{p,0} (N)$.  The solutions with $n >0$ have smaller condensation energy.   They are relevant to fluctuations at finite $T$, but  in the Eliashberg approximation, there is just one  transition temperature, and the gap function has no nodes along the Matsubara axis.   However, as  $\gamma$ increases, condensation energies of the solutions for $n >0$  get progressively closer to the one for $n=0$, and, finally,  the condensation energies  for all finite $n$ become the same as for $n=0$, while the ones for $n \to \infty$ form a continuous spectrum.    This happens for $\gamma \geq1$, when $N \neq 1$ and for $\gamma  =2$ for the original $\gamma$ model with $N=1$.    Once the spectrum becomes continuous,  a new channel for low-energy longitudinal  fluctuations emerges, with a number of consequences.     The key goal of the study of the gamma model is to demonstrate that it has this new physics.  However, to show this,  we first have to establish the presence of the  discrete set of solutions for smaller $\gamma$, and, most important, to  prove that the set is infinite and extends to $n=\infty$ (the solution of the linearized gap equation).   This is what we do in this paper (and in paper I), where we study the gamma model for smaller $\gamma <1$ and show how the infinite set of solutions emerges in the quantum critical, non-Fermi-liquid regime.

\section{The $\gamma$-model, Eliashberg equations}
\label{sec:model}

The $\gamma$ model was introduced in paper I and in earlier papers, and we refer the reader to these works for the justification of the model and its relation to various quantum-critical systems, e.g., fermions at the verge of a nematic of spin-density wave/charge-density wave transition, or Sachdev-Ye-Kitaev-type models of dispersion-less fermions randomly interacting with optical phonons.  The model describes low-energy fermions, interacting by exchanging soft bosonic excitations.
 The effective dynamical four-fermion interaction,
 which
  contributes to both the fermionic self-energy  $\Sigma (k_F, \omega_m) = \Sigma (\omega_m)$ and the pairing vertex $\Phi (\omega_m)$, is $V(\Omega_m) = {\bar g}^\gamma/|\Omega_m|^\gamma$ with the exponent $\gamma$. [For $\Phi (\omega)$,  this holds in an attractive channel with a particular spatial symmetry, determined by the underlying model with full momentum dependence of the interaction].
 We assume, like in earlier works, that soft bosons are slow compared to dressed fermions.  In this
situation, the coupled equations for  $\Sigma (\omega_m)$ and $\Phi (\omega_m)$ are similar to Eliashberg equations for the case when the effective dynamical interaction is mediated by phonons and we will use the term ``Eliashberg equations" for our case.

At a finite $T$ the coupled Eliashberg equations for  $\Phi (\omega_m)$ and $\Sigma (\omega_m)$ are,
 in Matsubara formalism
    \bea \label{eq:gapeq}
    \Phi (\omega_m) &=&
     {\bar g}^\gamma \pi T \sum_{m'} \frac{\Phi (\omega'_{m})}{\sqrt{{\tilde \Sigma}^2 (\omega'_{m}) +\Phi^2 (\omega'_{m})}}
    ~\frac{1}{|\omega_m - \omega'_{m}|^\gamma}, \nonumber \\
     {\tilde \Sigma} (\omega_m) &=& \omega_m
   +  {\bar g}^\gamma \pi T \sum_{m'}  \frac{{\tilde \Sigma}(\omega'_m)}{\sqrt{{\tilde \Sigma}^2 (\omega'_{m})  +\Phi^2 (\omega'_{m})}}
    ~\frac{1}{|\omega_m - \omega'_{m}|^\gamma}
\eea
 where ${\tilde \Sigma}(\omega_{m}) = \omega_m + \Sigma (\omega_m)$. Observe that we define $\Sigma (\omega_m)$ with the overall plus sign and  without the overall factor of $i$, that is $G^{-1} (k, \omega_m) = i {\tilde \Sigma} (\omega_m) - \epsilon_k$. In these notations, $\Sigma (\omega_{m})$ is a  real function, odd in frequency.

 The  superconducting gap function $\Delta (\omega_m)$ is defined as
\beq
 \Delta (\omega_m) = \omega_m  \frac{\Phi (\omega_m)}{{\tilde \Sigma} (\omega_m)} = \frac{\Phi (\omega_m)}{1 + \Sigma (\omega_m)/\omega_m}
  \label{ss_1}
  \eeq
   The equation for $\Delta (\omega_{m})$ is readily obtained from (\ref{eq:gapeq}):
   \beq
   \Delta (\omega_m) = {\bar g}^\gamma \pi T \sum_{m'} \frac{\Delta (\omega'_{m}) - \Delta (\omega_m) \frac{\omega'_{m}}{\omega_m}}{\sqrt{(\omega'_{m})^2 +\Delta^2 (\omega'_{m})}}
    ~\frac{1}{|\omega_m - \omega'_{m}|^\gamma}.
     \label{ss_11}
  \eeq
   This equation contains a single function $\Delta (\omega_{m})$, but at the cost that $\Delta (\omega_m)$ appears also in the right-hand-side (rhs).  Both $\Phi (\omega_m)$ and $\Delta (\omega_m)$ are defined up to an overall $U(1)$ phase factor, which we set to zero.

The full set of Eliashberg equations for electron-mediated pairing contains the additional equation, which describes the feedback from the pairing on $V(\Omega)$. In this paper
 we do not include this feedback into consideration, to avoid additional complications.  In general terms, the feedback from the pairing makes bosons less incoherent and can be modeled by taking  the exponent
  $\gamma$ to be $T$ dependent and by moving it towards larger values as $T$ decreases.

Both $\Phi (\omega_m)$ and $\Sigma (\omega_m)$ contain divergent contributions from the $m'=m$ terms in the summation over internal Matsubara frequencies. The divergencies, however, cancel out in the gap equation (\ref{ss_11}) by the Anderson theorem~\cite{agd}, because scattering with zero-frequency transfer mimics the effect of scattering by nonmagnetic impurities.
 To get rid of divergencies in (\ref{eq:gapeq}), one can use the same trick as in the derivation of the Anderson theorem:  pull out the terms with $m'=m$ from the frequency sums, move them  to the left-hand-side (lhs). of the equations, and introduce new variables  $\Phi^* (\omega_m)$ and $\Sigma^* (\omega_m)$ as
   \bea
   \Phi^* (\omega_m) & = & \Phi (\omega_m)
   \left(1- Q (\omega_m)\right), \nonumber \\
   {\tilde \Sigma}^* (\omega_m) &=& {\tilde \Sigma} (\omega_m) \left(1- Q (\omega_m)\right)
   \label{ss_2}
   \eea
    where
    \beq
     Q (\omega_m) =   \frac{\pi T V
     %_{\rm eff}
      (0)}{\sqrt{{\tilde \Sigma}^2 (\omega_{m}) +\Phi^2 (\omega_{m})}}
   \label{ss_2_a}
   \eeq
    The ratio $\Phi (\omega_m)/ {\tilde \Sigma} (\omega_m) = \Phi^* (\omega_m)/ {\tilde \Sigma}^* (\omega_m)$, hence $\Delta (\omega_m)$, defined in (\ref{ss_1}),  is invariant under this  transformation.
     One can easily verify that the equations on $\Phi^* (\omega_m)$ and ${\tilde \Sigma}^* (\omega_m)$  are the same as in (\ref{eq:gapeq}), but  the summation over $m'$ now excludes the divergent term with $m' =m$.
   To simplify the formulas, we re-define $\Phi^*$ and $\tilde{\Sigma}^*$ back as $\Phi$ and $\tilde{\Sigma}$.
    We have
  \bea \label{eq:gapeq_a}
    \Phi (\omega_m) &=&
     {\bar g}^\gamma \pi T \sum_{m' \neq m} \frac{\Phi (\omega'_{m})}{\sqrt{{\tilde \Sigma}^2 (\omega'_{m}) +\Phi^2 (\omega'_{m})}}
    ~\frac{1}{|\omega_m - \omega'_{m}|^\gamma}, \nonumber \\
     {\tilde \Sigma} (\omega_m) &=& \omega_m
   +  {\bar g}^\gamma \pi T \sum_{m' \neq m}  \frac{{\tilde \Sigma}(\omega'_m)}{\sqrt{{\tilde \Sigma}^2 (\omega'_{m})  +\Phi^2 (\omega'_{m})}}
    ~\frac{1}{|\omega_m - \omega'_{m}|^\gamma}
\eea

In the normal state ($\Phi \equiv 0$), the self-energy is ($m >0$)
  \beq
  \Sigma (\omega_m) =  {\bar g}^\gamma (2\pi T)^{1-\gamma} \sum_{m'=1}^m \frac{1}{|m'|^{\gamma}} = {\bar g}^\gamma (2\pi T)^{1-\gamma} H_{m, \gamma}
\label{ss_111_a}
\eeq
where $H_{m, \gamma}$ is the harmonic number.  This expression holds for $\omega_m \neq \pm \pi T$. For the two lowest  Matsubara frequencies, $\Sigma (\pm \pi T) =0$.  This will be essential for our paper.  We remind the reader that
 $\Sigma (\omega_m)$ in (\ref{ss_111_a}) is not the full self-energy, as the summation is only over
 $m' \neq m$
  The full self-energy includes also the term with
  $m' =m$
   and is nonzero for all  Matsubara frequencies.

We now extend the model and introduce a parameter $N$ to control the relative strength of the interactions in the particle-hole and particle-particle channels.
   In essence we just assume that the effective, momentum-integrated interactions in the two channels have
   the same frequency dependence, but different amplitudes.
   We treat $N$ as a continuous variable, but keep the same notation as in the original work, Ref. ~\cite{raghu_15}),  where  the original model was extended to  a matrix $SU(N)$ model, and $N$ was treated as an integer.
     With this extension
 \bea \label{eq:gapeq_b}
    \Phi (\omega_m) &=&
     \frac{{\bar g}^\gamma}{N} \pi T \sum_{m' \neq m} \frac{\Phi (\omega'_{m})}{\sqrt{{\tilde \Sigma}^2 (\omega'_{m}) +\Phi^2 (\omega'_{m})}}
    ~\frac{1}{|\omega_m - \omega'_{m}|^\gamma}, \nonumber \\
     {\tilde \Sigma} (\omega_m) &=& \omega_m
   +  {\bar g}^\gamma \pi T \sum_{m' \neq m}  \frac{{\tilde \Sigma}(\omega'_m)}{\sqrt{{\tilde \Sigma}^2 (\omega'_{m})  +\Phi^2 (\omega'_{m})}}
    ~\frac{1}{|\omega_m - \omega'_{m}|^\gamma}
\eea
and
  \beq
   \Delta (\omega_m) = \frac{{\bar g}^\gamma}{N} \pi T \sum_{m' \neq m} \frac{\Delta (\omega'_{m}) - N \Delta (\omega_m) \frac{\omega'_{m}}{\omega_m}}{\sqrt{(\omega'_{m})^2 +\Delta^2 (\omega'_{m})}}
    ~\frac{1}{|\omega_m - \omega'_{m}|^\gamma}.
     \label{ss_11_b}
  \eeq
 We emphasize that for our paper the extension to $N \neq 1$ is just a convenient way to better understand the interplay between the tendencies towards NFL and pairing. Our ultimate goal is to understand the physics in the physical case of $N=1$.  This is why we first eliminated the  divergent terms, which cancel out in the gap equation for $N=1$, and only then extended the model to $N \neq 1$.  An alternative approach, suggested in Ref. \cite{Wang_H_18}, is to extend to $N \neq 1$ without first subtracting the $m'=m$ terms in the equations for $\Phi(\omega_m)$ and $\Sigma (\omega_m)$. In this case, one has to deal with the actual divergencies in the rhs of these equations and also in the gap equation.
  Using the same analogy as before, once the interactions in the
   particle-hole and the particle-particle channel become nonequivalent (the case $N \neq 1$),  thermal fluctuations effectively act as singular magnetic impurities, which do not cancel out in  the gap equation~\cite{ag_imp}.

 We will chiefly analyze the linearized equation for the pairing vertex $\Phi (\omega_m)$
   \beq
    \label{eq:gapeq_c}
    N \Phi (\omega_m) =
     {\bar g}^\gamma \pi T \sum_{m' \neq m} \frac{\Phi (\omega'_{m})}{|{\tilde \Sigma} (\omega'_{m})|}
    ~\frac{1}{|\omega_m - \omega'_{m}|^\gamma}
    \eeq
    and the gap function
   \beq
    \label{eq:gapeq_d}
    \Delta (\omega_m) = \frac{{\bar g}^\gamma}{N} \pi T \sum_{m' \neq m} \frac{\Delta (\omega'_{m}) - N \Delta (\omega_m) \frac{\omega'_{m}}{\omega_m}}{|\omega'_{m}|}
    ~\frac{1}{|\omega_m - \omega'_{m}|^\gamma}.
    \eeq
    The two equations are equivalent in the sense that both have a
     nontrivial solution at the onset temperature for the pairing. We label this temperature as $T_p$ rather than superconducting $T_c$.
       because in general
         $T_c$ is smaller than $T_p$ due to gap fluctuations.  In this paper we focus on the solution of the Eliashberg equations and neglect gap fluctuations, in which case, $T_p = T_c$.
          The comparative analysis of $T_p$ and $T_c$ will be presented in a separate paper.

To distinguish between finite $T$ and $T=0$, below we label the gap function at a finite $T$
 as $\Delta_n (m)$, where $m$ is a discrete Matsubara number and $\omega_m = \pi T(2m+1)$, and at $T=0$
 as $\Delta (\omega_m)$, where $\omega_m$ is a continuous Matsubara frequency. We use the same notations for other quantities.

\section{The linearized gap equation, analytic consideration}
\label{sec:analytics}
We re-write the self-energy (\ref{ss_111_a}) as
\beq
\Sigma (m) = \pi T K A (m) {\text {sgn}} (2m +1)
\label{sigma}
\eeq
 where
 \bea
 A (m) &=& 2 \sum_{1}^{m} \frac{1}{n^\gamma} , ~~m >0 \nonumber \\
 A (m) &=& A (-m-1), ~~ m<-1 \nonumber\\
 A (0) &=& A (-1) =0
 \eea
and
\beq
K = \left(\frac{{\bar g}}{2\pi T}\right)^\gamma
\eeq
Using these notations and the fact that $\Phi (m)$ is even under $m \to -m-1$,
  we re-write
 Eq. (\ref{eq:gapeq_c}) for $\Phi (m)$ as the set of two coupled equations, by singling out $\Phi (0) = \Phi (-1)$ (Ref. \cite{Wang2016}):
 \begin{widetext}
 \bea
 &&
 (N-K)\Phi (0)  = \sum_{n=1}^\infty \frac{\Phi (n)}{A (n) + \frac{2n+1}{K}} \left(\frac{1}{n^\gamma} + \frac{1}{(n+1)^\gamma}\right) \label{ch_1} \\
 && \Phi (m>0) = \frac{1}{N} \sum_{n=1,n\neq m}^\infty \frac{\Phi (n)}{A (n) + \frac{2n+1}{K}} \left(\frac{1}{|n-m|^\gamma} + \frac{1}{(n+m+1)^\gamma}\right) + \frac{\Phi (0) K}{N} \left(\frac{1}{n^\gamma} + \frac{1}{(n+1)^\gamma}\right) \label{ch_2}
 \eea
 Eliminating $\Phi (0)$ from this set we obtain
 \bea
 N \Phi (m>0) &=&  \sum_{n=1,n\neq m}^\infty \frac{\Phi (n)}{A (n) + \frac{2n+1}{K}} \frac{1}{|n-m|^\gamma} + \sum_{n=1}^\infty \frac{\Phi (n)}{A (n) + \frac{2n+1}{K}}\frac{1}{(n+m+1)^\gamma} + \nonumber \\
  && \frac{K}{N-K}  \sum_{n=1}^\infty \frac{\Phi (n)}{A (n) + \frac{2n+1}{K}} \left(\frac{1}{n^\gamma} + \frac{1}{(n+1)^\gamma}\right) \left(\frac{1}{m^\gamma} + \frac{1}{(m+1)^\gamma}\right)
   \label{ch_3}
 \eea
\end{widetext}
  A quick inspection of  Eq. (\ref{ch_3}) shows that at low $T$, when $K \gg 1$, there are two regions of $N$,
    in which the solution with $\Phi (m) \neq 0$ may appear: a) $N \approx K$, such that $N -K = {\text {const}}$, and b) $N = O(1)$, such that $K\gg N$. We consider these two regions separately.

\subsection{The region of large $N$,   $N \approx K$}
\label{sec:Nlarge}

\begin{figure}
  \includegraphics[width=8cm]{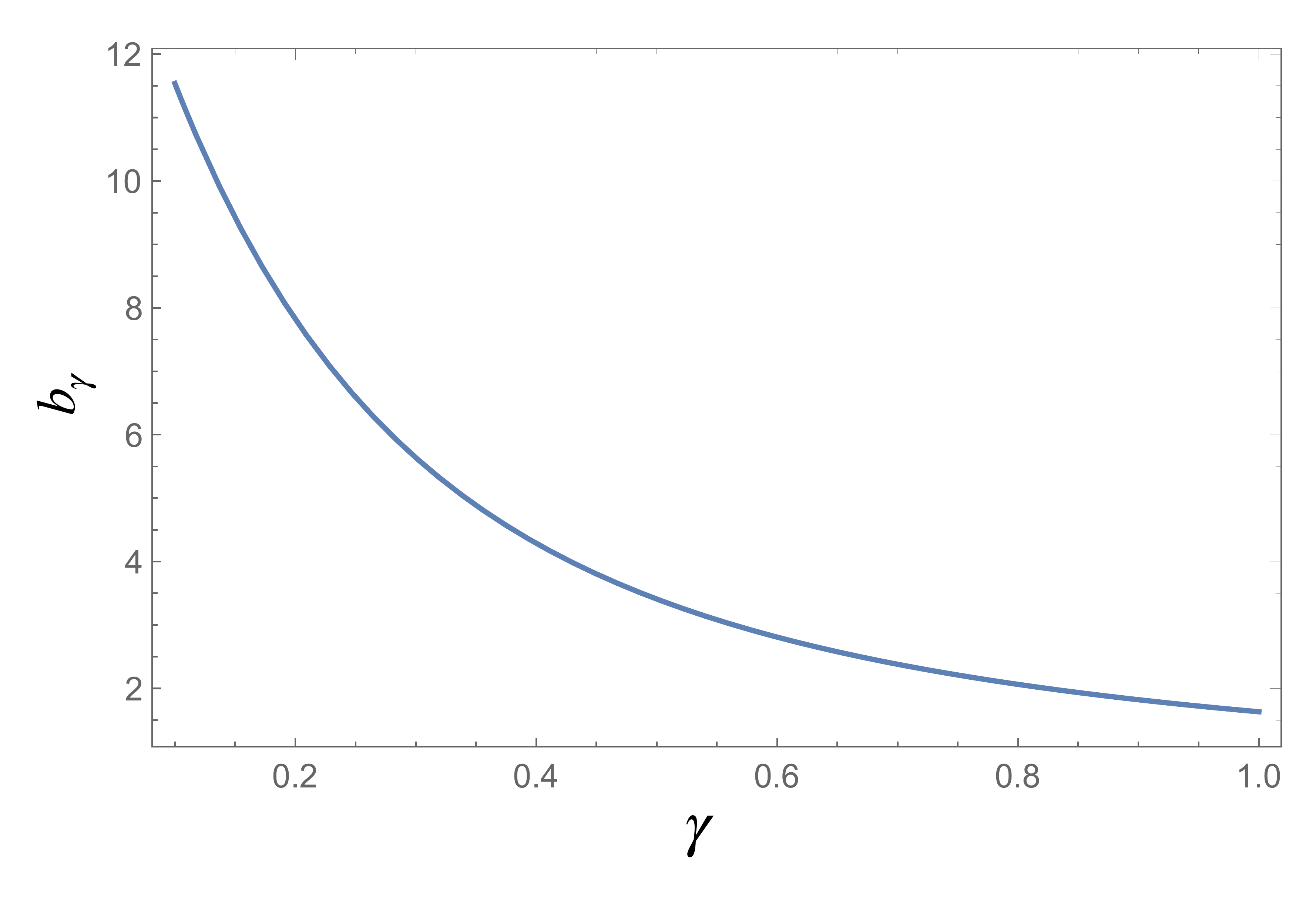}
  \caption{The plot of $b_\gamma$ vs $\gamma$, Eq.\eqref{ch_6}.}\label{fig:bgamma}
\end{figure}

We introduce $b_\gamma$ via $N$ as $N = K + b_\gamma$, substitute this relation into (\ref{ch_3}), and take the limit $K \to \infty$.  The divergent $K$ cancels out, and we obtain
\beq
b_\gamma \Phi (m>0) =   \sum_{n=1}^\infty \frac{\Phi (n)}{A (n)}
 \left(\frac{1}{n^\gamma} + \frac{1}{(n+1)^\gamma}\right) \left(\frac{1}{m^\gamma} + \frac{1}{(m+1)^\gamma}\right) \label{ch_4}
 \eeq
The form of the kernel implies that the solution should be in the form
\beq
\Phi (m>0) =   C  \left(\frac{1}{m^\gamma} + \frac{1}{(m+1)^\gamma}\right)
 \label{ch_5}
 \eeq
 Substituting this form into (\ref{ch_4}) we obtain that  $b_\gamma$ is given by
  \beq
 b_\gamma =  \sum_{n=1}^\infty \frac{1}{A (n)}
 \left(\frac{1}{n^\gamma} + \frac{1}{(n+1)^\gamma}\right)^2
  \label{ch_6}
 \eeq
  At large $n$, $A (n) \approx 2 n^{1-\gamma}/(1-\gamma)$.  The sum in the rhs of (\ref{ch_6}) then converges for all $0< \gamma <1$,  i.e., $b_\gamma$ is indeed of order 1.  We plot $b_\gamma$ in Fig.\ref{fig:bgamma} At small $\gamma$, $b_\gamma \propto 1/\gamma$. At $\gamma =1$, $b_1 =1.63303$.

For the onset temperature for the pairing  we obtain
 \beq
T_p = \left(\frac{{\bar g}}{2\pi}\right) \frac{1}{N^{1/\gamma}} \left(1 + \frac{b_\gamma}{N\gamma}\right)
\label{ch_7}
\eeq
We emphasize that $T_{p}$ remains finite  for arbitrary large $N$, i.e., for arbitrary weak pairing interaction.  Eq. (\ref{ch_7}) has been earlier obtained in Ref. \cite{Wang2016} using somewhat different computational procedure.

Using (\ref{ch_1}), (\ref{ch_5}), and (\ref{ch_6}), we find $\Phi (0) = C$. Comparing with (\ref{ch_5}), we see that the magnitude of the pairing vertex at Matsubara frequencies $\omega_m = \pm \pi T$ is of the same order as at other frequencies [a more detailed analysis shows that $\Phi (m)$ is the largest at $m = 2$, see Fig. \ref{fig_T_2}].  At the same time, the pairing gap $\Delta (m)$ is strongly peaked at $m=0,-1$:  $\Delta (0) = \Delta (-1) = C$, while $\Delta (m>0) =\Delta(0)/K$ is much smaller.  In fact, the leading term in the expression for $T_{p}$ in Eq. (\ref{ch_7}) can be obtained  from the gap equation (\ref{ss_11_b}) if we keep there only the terms with $m =0,-1$ and use the fact that $\sum_m' {\text{sign}} (2m'+1)/|m'|^\gamma =0$. In this case, the gap equation simplifies to
\bea
\Delta (0) &=& \frac{{\bar g}^\gamma}{N} \frac{\Delta (-1)}{(2\pi T)^\gamma} \nonumber \\
\Delta (-1) &=& \frac{{\bar g}^\gamma}{N} \frac{\Delta (0)}{(2\pi T)^\gamma}
\label{ch_8}
\eea
Solving this set we obtain $T_{p} = {\bar g}/(2\pi N^{1/\gamma})$.
 The outcome is that at large $N$, the pairing predominantly involves fermions with Matsubara frequencies $\omega_m = \pm \pi T$ (Matsubara numbers $m=0,-1$), for which $\Sigma (\omega_m)$ in (\ref{sigma}) vanishes.
  The gap function at other Matsubara frequencies is induced by a nonzero $\Delta (0)= \Delta (-1)$.

Below we classify the solutions for the pairing vertex function by an integer $n$, which is the number of times $\Phi_m$ changes sign as a function of Matsubara number $m$. In this classification, the sign-preserving pairing vertex $\Phi (m)$  in (\ref{ch_5}) corresponds to $n=0$. Accordingly, $T_p$ in (\ref{ch_7}) is $T_{p,0}$ and $\Phi (m) = \Phi_0 (m)$, $\Delta (m) = \Delta_0 (m)$.

\begin{figure}
 	\includegraphics[width=12cm]{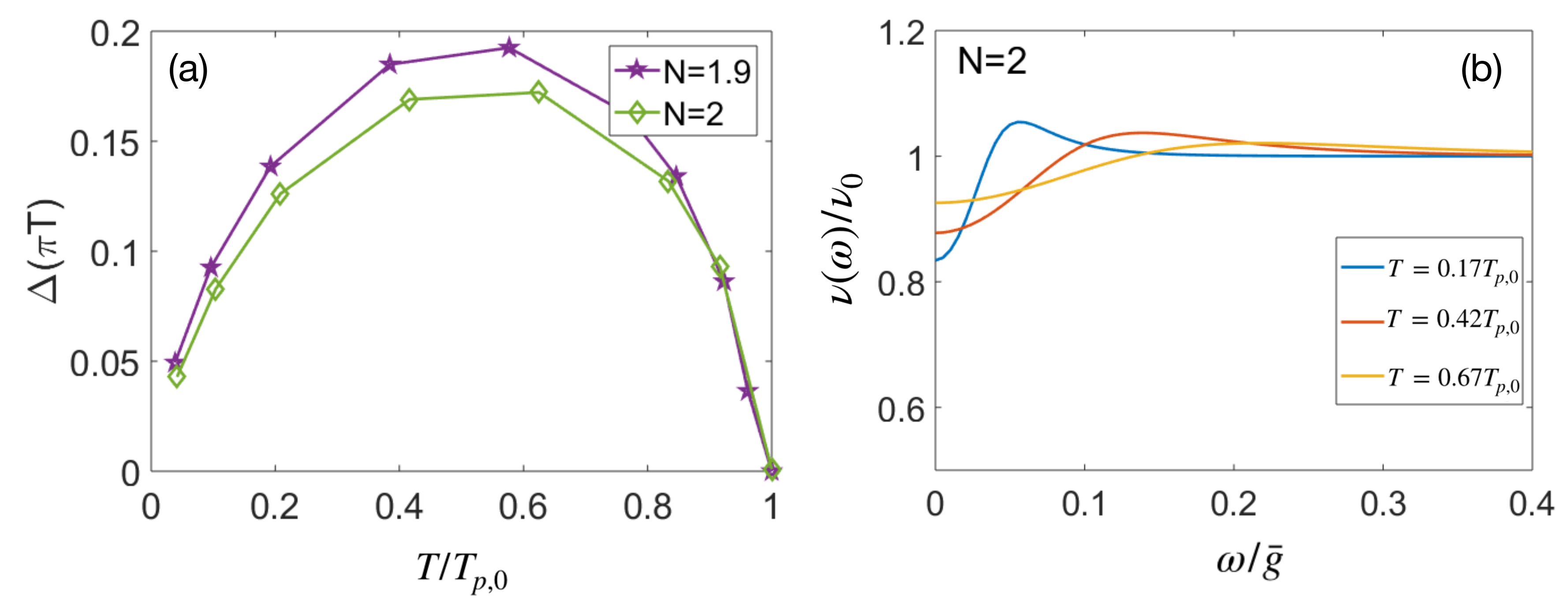}
 	\caption{The gap function and the density of states below $T_{p,0}$, for $N > N_{cr}$. We set
   $\gamma=0.9$, in which case $N_{cr} = 1.340$.
  Panel (a): $\Delta (m=0) \equiv \Delta(\pi T)$ as a function of $T$.  Panel (b): the density of states $\nu(\omega)$ in real frequency, normalized to its value in the normal state.  The density of states shows a gap-filling behavior: as $T$ increases towards $T_{p,0}$, states at low frequencies fill in, and the position of the maximum shifts to higher $\omega$
   }\label{fig:delta0}
 \end{figure}

At a first glance, the result that $T_{p,0}$ is nonzero  at large $N$  contradicts the
 $T=0$ analysis in paper I and earlier studies~\cite{Wang2016,Wang_H_17,Wang_H_18,Wu_19_1,Abanov_19,Chubukov_2020a}
that at $T=0$ there exists critical $N =N_{cr}$, which separates the state with a finite $\Delta (\omega_m)$ for $N < N_{cr}$ and the NFL normal state for $N > N_{cr}$. We, however,  showed in Refs. \cite{Wu_19_1,Abanov_19} that $\Phi (m)$ and $\Delta (m)$  evolve nonmonotonically below $T_{p,0}$ and for $N > N_{cr}$  vanish at $T \to 0$. We show the behavior of $\Delta (m=0) \equiv \Delta (\pi T)$ vs temperature in Fig.\ref{fig:delta0}a.  Such a state
 has a number of other unusual properties, e.g.,
 the gap-filling behavior of the density of states and the spectral function (see Fig.\ref{fig:delta0}b).   We refer the reader to Refs. \cite{Wu_19_1,Abanov_19,Chubukov_2020a} for the detailed analysis of the Eliashberg equations below $T_{p,0}$ and of the role of gap fluctuations.
 In Appendix~\ref{appendix_1} we compute the  free energy $F_{0,p}$ of a state with a finite $\Delta_0 (m)$  for $N > N_{cr}$ and analyze  $\delta F = F_{p,0} - F_n$, where $F_n$ is the free energy at $\Delta_0 (m) =0$.
We use $\delta F = \delta E - T \delta S$ and study the two terms separately.
 We find that at  $ T \leq T_{p,0}$, negative $\delta F$ comes primarily from negative $\delta E$, while
  $- T \delta S$ is positive. However, below a certain $T$, close to where $\Delta (m=0)$ changes its behavior in Fig. \ref{fig:delta0},  $-T\delta S$ becomes negative, and $\delta E$ almost vanishes.  In this $T$ range, a negative $\delta F$ comes primarily from the entropy.  This is consistent  with the
   vanishing of $\Delta_0 (m)$ at $T=0$.

\begin{figure}
  \includegraphics[width=15cm]{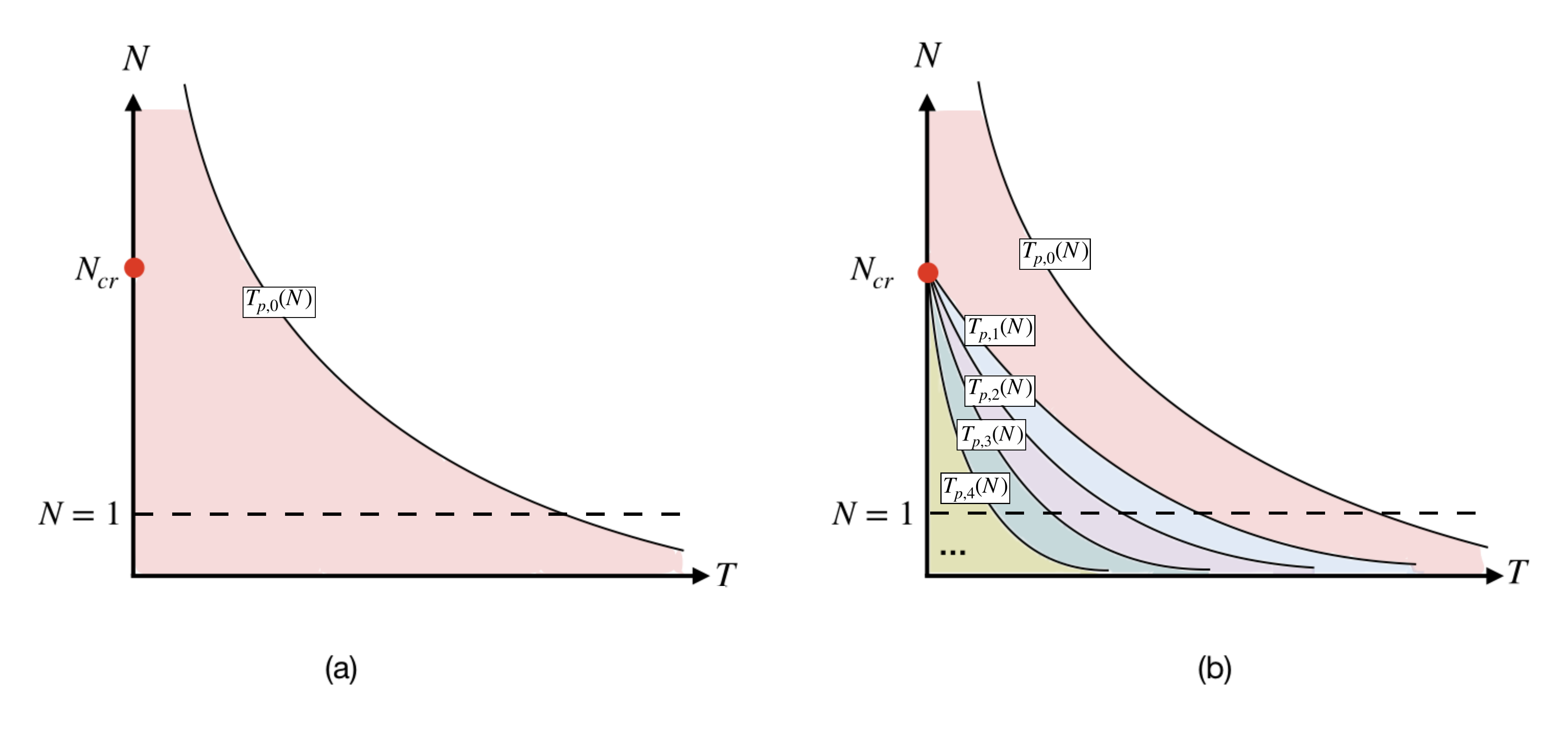}
  \caption{Two potential phase diagrams of the $\gamma$ model.
 (a) There exists only one onset temperature $T_{p,0}$ for any given $N$, like in BCS theory.
   In this case, $N_{cr}$ is an isolated  $T=0$ QCP, with no transition line attached to it.
   (b) There exists an  infinite set of $T_{p,n}$, which all terminate at
     $N=N_{cr}$. In this case, a QCP at $N=N_{cr}$ is critical point of an infinite order.  We show that the
      correct phase diagram is the one in panel (b).   }\label{fig:illustration}
\end{figure}

Now, if $T_{p,0}$ was the only solution of the linearized gap equation,  the phase diagram
 within the Eliashberg theory would be as in Fig.\ref{fig:illustration}(a), i.e., there would be an isolated QCP at $T=0$ and $N= N_{cr}$, with no transition line coming out of it.
 We show below that this is not the case, and the actual phase diagram is the one in  Fig.\ref{fig:illustration}(b), with an infinite number of  lines terminating at $N_{cr}$.
 For this to hold,  the linearized gap equation must have an infinite set of onset temperatures $T_{p,n}$
  for any $N < N_{cr}$. This is what we analyze next.

\subsection{The region $N = O(1)$  .}
\label{sec:N_1}

Our reasoning to search for additional solutions of the gap equation in the region $N = O(1)$  comes from the $T=0$ analysis in  1aper I.
  The key result of paper I is that for any $N < N_{cr}$ [which is $O(1)$ for a generic $\gamma <1$],
   there exists an infinite, discrete set of  solutions of the nonlinear gap equation, $\Delta_n (\omega_m)$. If system properties evolve smoothly between $T=0$ and $T >0$, each gap function evolves with $T$ and has to vanish at  some finite $T_{p,n}$. By this logic,  there must be an infinite set of  $T_{p,n}$ for any given $N < N_{cr}$.  Because all $\Delta _n (\omega_m)$  at $T=0$ vanish at $N= N_{cr}$,   all $T_{p,n} (N)$ with finite $n$ must approach $0$ at $N=N_{cr}$, as in Fig.\ref{fig:illustration}(b).

  We first verify the very assumption that the system behavior evolves smoothly between $T=0$ and $T >0$. This is not a'priori guaranteed, as at $T=0$  the infinite set of solutions for all $N < N_{cr}$ emerges due to a fine balance between the tendencies towards NFL and pairing.  A  finite $T$ perturbs this balance, and it is
  possible in principle that an infinite set of solutions exists only at $T=0$, while at a finite $T$ only a single solution with a finite $\Delta$ survives.

To address this issue, we recall that  the origin for the appearance of the set of $\Delta_n (\omega_m)$  at $T=0$
 is the existence of the solution of the linearized gap equation for all $N < N_{cr}$ rather than only for $N = N_{cr}$.  In paper I we  obtained the exact solution of the linearized gap equation for all $N < N_{cr}$ and $0<\gamma <1$.
 At small $\omega_m \ll {\bar g}$, the exact solution  reduces to
 \beq
 \Delta (\omega_m) =  E |\omega_m|^{\gamma/2} \cos{\left(\beta_N \log{\frac{|\omega_m|}{{\bar g}}} + \phi_N\right)}
\label{ch_15}
\eeq
where $E$ is an infinitesimally small overall magnitude,  $\phi_N = O(1)$ is the phase factor, and $\beta_N$ is the real solution of $\epsilon_{\beta_N} =N$, where
\beq
\epsilon_\beta = \frac{1-\gamma}{2}\frac{|\Gamma (\gamma /2(1 +2i\beta) )|^{2}}{\Gamma (\gamma )}\left(1+\frac{\cosh (\pi \gamma \beta)}{\cos (\pi \gamma /2)} \right)
    \label{su_15_2}
\eeq
 A solution of this equation with a real $\beta = \pm \beta_N$  exists for $N <N_{cr}$, where
\beq
N_{cr} =  \frac{1-\gamma}{2}\frac{
\Gamma^2 (\gamma /2)}{\Gamma (\gamma )}
\left(1+\frac{1}{\cos (\pi \gamma /2)} \right)
\label{ch_16}
\eeq
One can verify that $N_{cr} >1$ for $0<\gamma <1$. At $\gamma \ll 1$, $N_{cr} \approx 4/\gamma$. At $\gamma \to 1$, $N_{cr} \to 1$.

\begin{figure}
  \includegraphics[width=8cm]{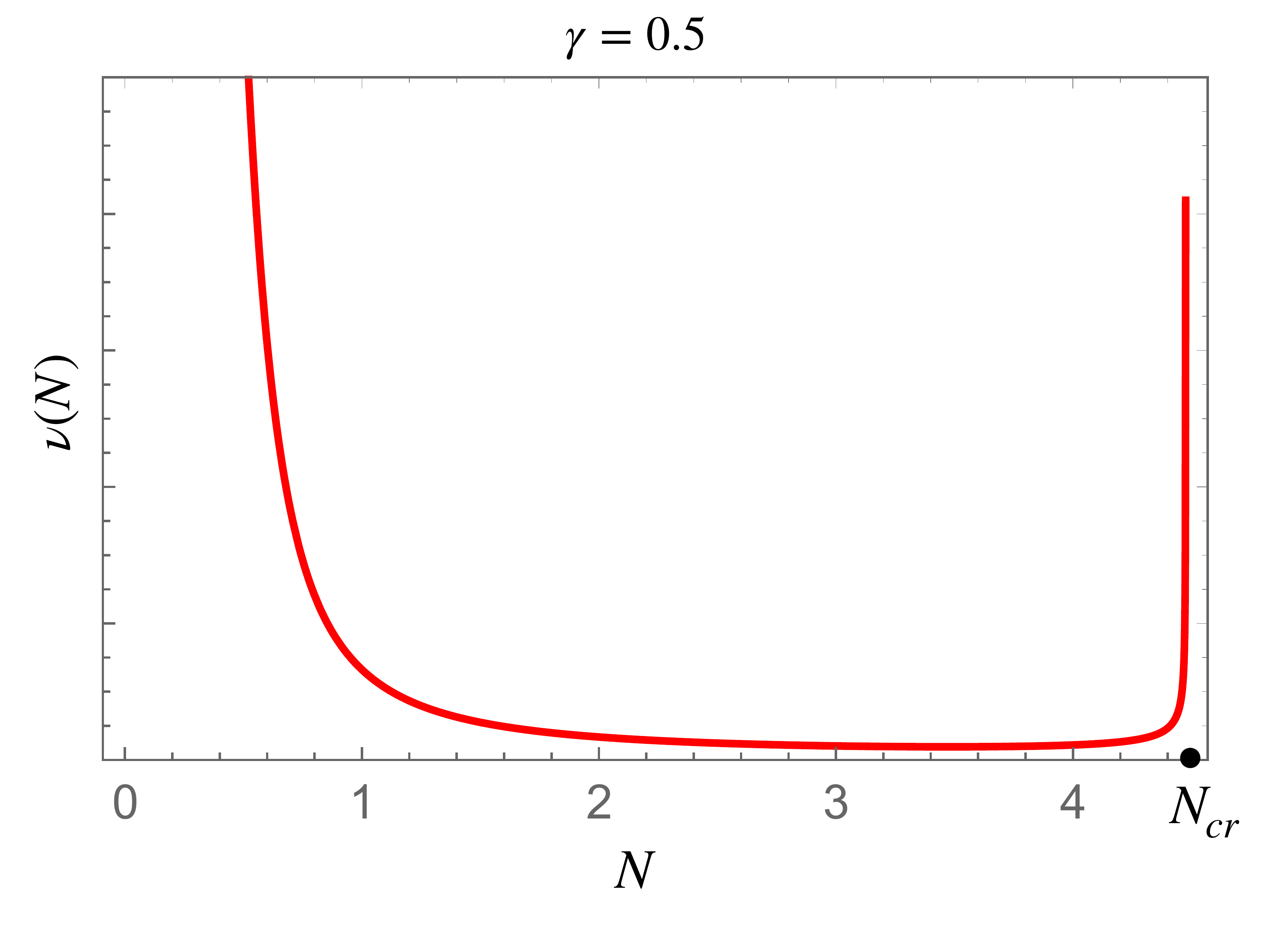}
  \caption{The DoE $\nu(N)$ (in arbitrary units) at $T=0$,  for $\gamma=0.5$. The DoE is non-zero for all $N < N_{cr}$. It has a strong singularity near $N=0$ and a weaker singularity at $N=N_{cr}$.
   }\label{fig:DOS}
\end{figure}

It is instructive to interpret $\epsilon_{\beta_N} =N$ as the dispersion relation and identify $\beta_N$ with the effective momentum and $N$ with the effective energy.  Then one can define the DoE as
\beq
\nu (N) \propto \left.\frac{d\beta}{d\epsilon_{\beta }} \right|_{\beta = \beta_N}.
\label{chh_11}
\eeq
  We plot this function in Fig.\ref{fig:DOS}.  As expected, it is nonzero for all $N < N_{cr}$.
   It is singular near
   $N=0$, where $\nu (N) \propto (1/N)^{(2-\gamma)/(1-\gamma)}$ and near
      $N_{cr}$, where $\nu (N) \propto 1/\sqrt{N_{cr}-N}$.  This last singularity, however, affects $\nu (N)$ only in the immediate vicinity of $N = N_{cr}$, as one can see from Fig. \ref{fig:DOS}.
 The DoE in (\ref{chh_11}) is defined up to an overall factor. In Fig.\ref{fig:DOS} we plot $\nu (N)$
without  normalizing it.

Now, for a smooth evolution of system properties between $T=0$ and $T >0$ the same $\nu (N)$ must emerge
if we solve the linearized gap equation by approaching the $T=0$ limit from a finite $T$.  To verify whether this is the case,  we keep $N = O(1)$ and set $K \propto 1/T^\gamma$ to infinity in (\ref{ch_3}).

\subsubsection{The limit $T \to 0$}

Keeping $N$ finite and setting $K$ to infinity, we obtain from (\ref{ch_3}),
  after symmetrization and rescaling
 \beq
 {\tilde \Phi}(m>0) =  \sum_{n=1,n\neq m}^\infty {\tilde \Phi}(n) \frac{K_{n,m}}{(B_m B_n)^{1/2}},
  \label{ch_9}
 \eeq
 where ${\tilde \Phi}(m) = \Phi(m) (S_m/A(m))^{1/2}$, $B_m = S_m A(m)$, and $S_m$ and $K_{n,m}$ are given by
 \beq
 S_{m} = N - \frac{1}{A(m)} \left(\frac{1}{(2m+1)^\gamma} - \left(\frac{1}{m^\gamma} + \frac{1}{(m+1)^\gamma}\right)^2\right),
 \label{ch_10}
 \eeq
 \begin{widetext}
 \beq
 K_{n,m} = \frac{1}{|n-m|^\gamma} + \frac{1}{(n+m+1)^\gamma} - \left(\frac{1}{m^\gamma} + \frac{1}{(m+1)^\gamma}\right)  \left(\frac{1}{n^\gamma} + \frac{1}{(n+1)^\gamma}\right)
 \label{ch_11}
 \eeq
 In explicit form,
\beq
 B_m = 2N \sum_{n=1}^\infty \frac{1}{n^\gamma}-  \frac{1}{(2m+1)^\gamma} + \left(\frac{1}{m^\gamma} + \frac{1}{(m+1)^\gamma}\right)^2
 \label{ch_12}
 \eeq
\end{widetext}

 At large $n,m \gg 1$, $B_m \approx 2N m^{1-\gamma}/(1-\gamma)$, $ K_{n,m} \approx \frac{1}{|n-m|^\gamma} + \frac{1}{(n+m)^\gamma}$, ${\tilde \Phi}(m) \approx \Phi(m)$, and the equation for the pairing vertex can be approximated by an integral equation
 \beq
 \Phi(m) = \frac{1-\gamma}{2N} \int_0^\infty dn \frac{\Phi(n)}{n^{1-\gamma}} \left(\frac{1}{|n-m|^\gamma} + \frac{1}{(n+m)^\gamma}\right)
 \label{ch_14}
 \eeq
 This equation has been analyzed in paper I and in earlier works~\cite{acf,*acs,*moon_2,Wang2016,raghu_15}. The solution is
 \beq
 \Phi(m>0) = \frac{E}{m^{\gamma/2}} \cos{\left(\beta_N \log{m} + \phi\right)}
\label{ch_15_a}
\eeq
where $\beta_N$ is the same as in (\ref{ch_15}) and $E$ is an infinitesimally small overall factor.
 For these $m$, $\Sigma_m \propto m^{1-\gamma}$, hence
 \beq
 \Delta(m>0) = E m^{\gamma/2} \cos{\left(\beta_N \log{m} + \phi\right)}
\label{ch_15_b}
\eeq
The functional form of $\Delta(m)$ is the same as for the $T=0$ solution, but at this stage $\phi$ is a free parameter.
 To be the solution of the gap equation for all $m$, this  $\Delta(m)$ (or $\Phi(m)$) has to match with the solution at small $m$, when the discreteness of Matsubara numbers becomes relevant.  If this can be achieved by fixing the
 value of $\phi$, then $\nu (N)$ at $T \to 0$ is the same as at $T =0$.

\begin{figure}
    \includegraphics[width=15cm]{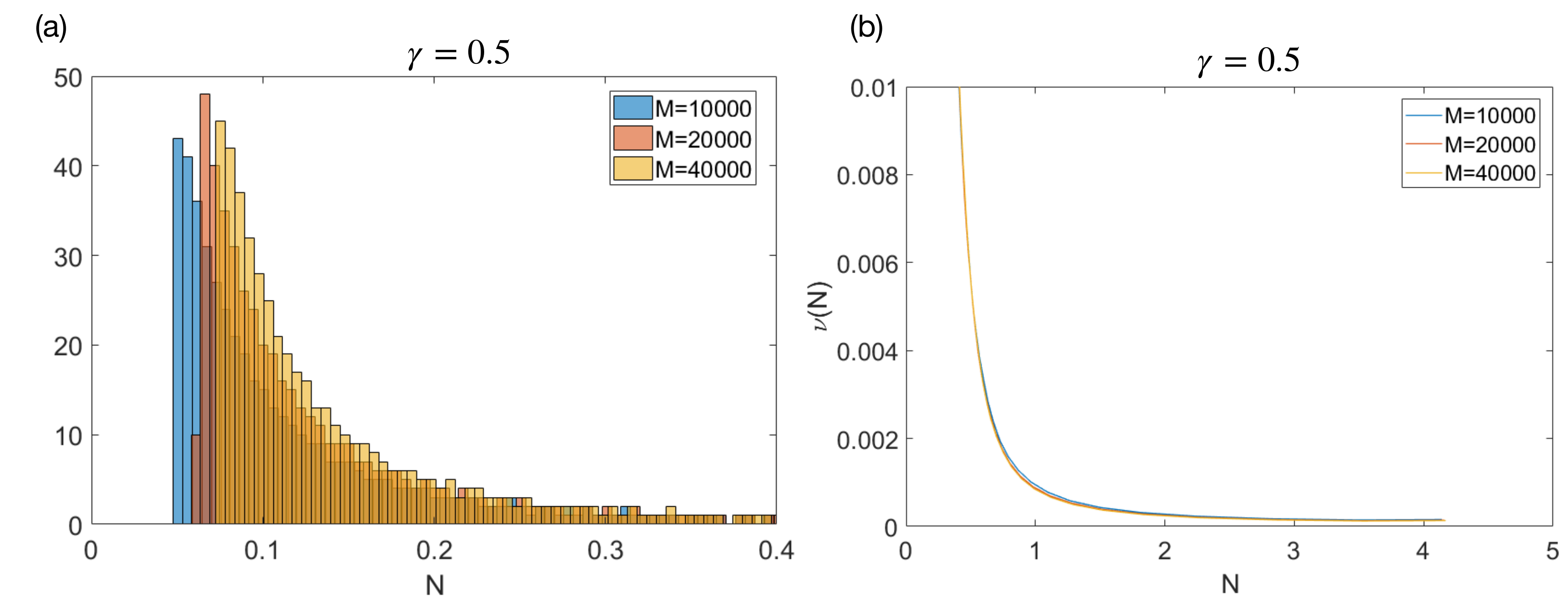}
    \caption{Numerical results for the  DoE at $T \to 0$.
     (a) The histogram of the eigenvalues for different numbers of sampled Matsubara points $M$. As $M$ increases, more eigenvalues are shifted towards larger $N$, and the number of eigenvalues in any given interval of $N$ increases. (b) the ``smoothened" DoE for different $M$, normalized such that $\nu(N=0.5)$ is kept fixed.
     As $M$ increases, $\nu (N)$ does not change much. The curves are similar to that in Fig. \ref{fig:DOS} except for the narrow peak near $N_{cr}$, as a much larger $M$ is required to get enough eigenvalues near $N_{cr}$. }\label{fig:DOShistogram}
  \end{figure}

This is similar to the $T=0$ case, where the oscillating $\Delta (\omega_m)$ from (\ref{ch_15}) has to match with the nonoscillating solution $\Delta (\omega_m) \propto 1/|\omega_m|^\gamma$ for $|\omega_m| \gg {\bar g}$. For $T=0$, the exact solution and the approximate, but highly accurate, analytical solution show that this is the case. Namely, for some particular $\phi = \phi_N$, the exact solution has the form of Eq. (\ref{ch_15}) for $|\omega_m| \ll {\bar g}$ and decays as $1/|\omega_m|^\gamma$ for $|\omega_m| \gg {\bar g}$.
 For $T \to 0$, we solve the gap equation numerically.   We present the results for the DoE  in Fig.\ref{fig:DOShistogram}.
 While for any finite number $M$ of Matsubara  points in numerical calculations the DoE consists of a discrete set of points, we see
     from the histogram of the eigenvalues in panel (a) that at  larger $M$, more eigenvalues move to larger $N$ and the number of eigenvalues in any fixed interval of $N$ increases.
   The  ``smoothened" $\nu (N)$ in panel (b)
weakly depends on $M$ and is quite similar to the DoE at $T=0$ in
      Fig. \ref{fig:DOS}.
     This strongly indicates  that   Eq. (\ref{ch_9}) has a solution for any $N < N_{cr}$, like at $T=0$,  i.e.  the system evolves continuously between $T=0$ and $ T \to 0$.
     %\addYW{(YW: Does the evolution of the histogram as a function of $M$ contain physical information, or merely represent the improvement of the numerics? Will we be able to make our point by showing the $M=40000$ histogram and its smoothened form only? From Fig. 5b I understand that the smoothened DoS barely changes from $M=10000$ to $M=40000$, while in Fig. 5a the histograms are visually different -- I guess this is simply because a $40000\times 40000$ matrix has 4 times as many eigennvalues than a $10000\times 10000$ one?)}

    %Wu: I think here we want to show that as M increases the DoE curve becomes closer to T=0 analysis. In fact normalization matters here and what we do is to keep DoE at some value, say N=0.5, fixed for different M, and that's why the curves look almost the same despite of various M. I we don't utilize this way, we may see the curve drops as we increase M, but this is not what we want.

\subsubsection{The case of small but finite  $T$}

We next consider the case when $T$ is small, but finite.
Like we said at the beginning of Sec. \ref{sec:N_1}, at $T=0$ and $N < N_{cr}$, there exists an infinite, discrete set of  solutions of the nonlinear gap equation, $\Delta_n (\omega_m)$. It is natural to expect that  each $\Delta_n$  smoothly evolves with $T$ and ends at a finite $T_{p,n}$. We then expect that  Eq. (\ref{ch_3}) should have an infinite set of solutions at large but finite $K$.  The equation for the pairing vertex in this case has the same form as in Eq. (\ref{ch_9}), but
  one should keep the bare $\omega$ term along with the self-energy, i.e.,  replace $A(m)$ by
  \beq
  A(m) + \frac{2m +1}{K}
  \label{ch_16_a}
  \eeq
  The implication here is that for large enough $m \sim K^{1/\gamma}$, the second term becomes comparable to the first one, and this will modify the functional form of $\Delta(m)$ compared to that in Eq. (\ref{ch_15_a}).
   The other dependence on $K$, from
    $K/(N-K) \approx  -1 +N/K$, is irrelevant as the $1/K$ term never becomes comparable to other terms in the rhs of (\ref{ch_3}).

 A qualitative argument for the existence of the infinite set of $T_{p,n}$ is the following:
 \begin{itemize}
 \item
  at $m \gg 1$, the difference between summation over Matsubara numbers $n$ and integration is negligible.
   Hence, the solution of the gap equation, expressed in terms of $\omega_m \approx 2 \pi m T$, should be the same as the exact solution at $T=0$, Eq. (\ref{ch_15}). In terms of $m$ we have
   \beq
 \Delta(m) = E m^{\gamma/2} \cos{\left(\beta_N \log{m} -  \beta_N \log{K^{1/\gamma}} + \phi_N\right)}
\label{ch_15_c}
\eeq
 The phase $\phi_N$ is fixed by the requirement that $\Delta(m) \propto 1/|m|^\gamma$ for $m > K^{1/\gamma}$, but there is another phase factor $-\beta_N \log{K^{1/\gamma}}$, which depends on $K$.
 \item
  From the analysis at $T \to 0$ we know that the gap equation has solutions for $N < N_{cr}$.
   The form of a solution is the same as  in (\ref{ch_15_c}), $\Delta(m) = E m^{\gamma/2}
    \cos{\left(\beta_N \log{m} +{\tilde \phi}_N \right)}$, with some particular
     ${\tilde \phi}_N$ which depends on $N$ (for a fixed  $\gamma$).
  \item
   We can  match the two forms of $\Delta(m)$ by relating the phases:
   \beq
   {\tilde \phi}_N = \phi_N - \beta_N \log{K^{1/\gamma}} + \pi n,
    \label{ch_17}
    \eeq
    where $n$ is an integer. Solving (\ref{ch_17}), we obtain the set of $K_n$, for which this identity holds.  Note that the additional factor is $\pi n$, not $2\pi n$, because we can independently  flip the sign of $\Delta(m) $ at $T \to 0$ keeping  $\Delta(m)$ at $T =0$ intact.
    \end{itemize}

  Solving  Eq. (\ref{ch_17}) for the critical temperature, we obtain
  \beq
  T_{p,n} \sim {\bar g} e^{-A n}, ~~ A = \frac{\pi}{\beta_N \gamma}
  \label{ch_18}
  \eeq
   This is consistent with the result in paper I that at $T=0$, $\Delta_n (\omega_m=0) \sim  {\bar g} e^{-A n}$.  As $N$ increases towards $N_{cr}$, $\beta_N \propto (N_{cr}-N)^{1/2}$ gets smaller, and
   all $T_{p,n}$ with $n >0$ become exponentially small in $N_{cr}-N$:
   \beq
   T_{p,n} \sim {\bar g} e^{-b n/(N_{cr}-N)^{1/2}},
   \label{chh_10}
   \eeq
  where $b = O(1)$. Eq. (\ref{chh_10}) shows that all $T_{p,n} (N)$ terminate simultaneously at $N = N_{cr}$.

    A remark is in order here.
     Eq. (\ref{ch_18}) and the $T=0$ result for $\Delta(m)$ are based on the assumption that the solution of the linearized gap equation oscillates up to $\omega_m \sim {\bar g}$ and decays as $1/|\omega_m|^\gamma$ at larger $\omega_m$. This  holds for most of $\gamma <1$, but for very small $\gamma$, oscillations extend to larger scale. In this case, Eq. (\ref{ch_17}) has to be modified. We discuss this in Appendix ~\ref{appendix_2}.

In Fig. \ref{fig_T}
 we show the numerical results for $T_{p,n}$ for $n =1-4$,
  along with the result for $T_{p,0}$ from Eq. (\ref{ch_7}).  We set representative $\gamma$ to be
  $\gamma =0.3$ and $\gamma =0.5$.  We verified that $T_{p,0}$ behaves as $1/N^{1/\gamma}$, as expected.
    At other $T_{p,n}$ lines exponentially approach zero as $N$ tends to $N_{cr}$, and the slope becomes larger as $n$ increases.  To obtain this behavior
       we used a
        ``hybrid frequency scale" method, which allowed us to numerically cover an exponentially large frequency range  and reach very low $T$, while keeping track of the Matsubara summation at the lowest frequencies. This is achieved by adopting a frequency mesh that overlaps with Matsubara frequencies $\omega_m=\pi T, 3\pi T, \cdots$ at small values and crosses over to a logarithmical spacing beyond a certain scale, above which the discreteness of the Matsubara sum becomes unimportant.
                 We discuss this method in Appendix~\ref{app:A}. It was also used in Ref.~\onlinecite{Wang_H_18} in addressing the interplay between the first-Matsubara physics and thermal fluctuations.

\begin{figure*}
	\begin{center}
		\includegraphics[width=13cm]{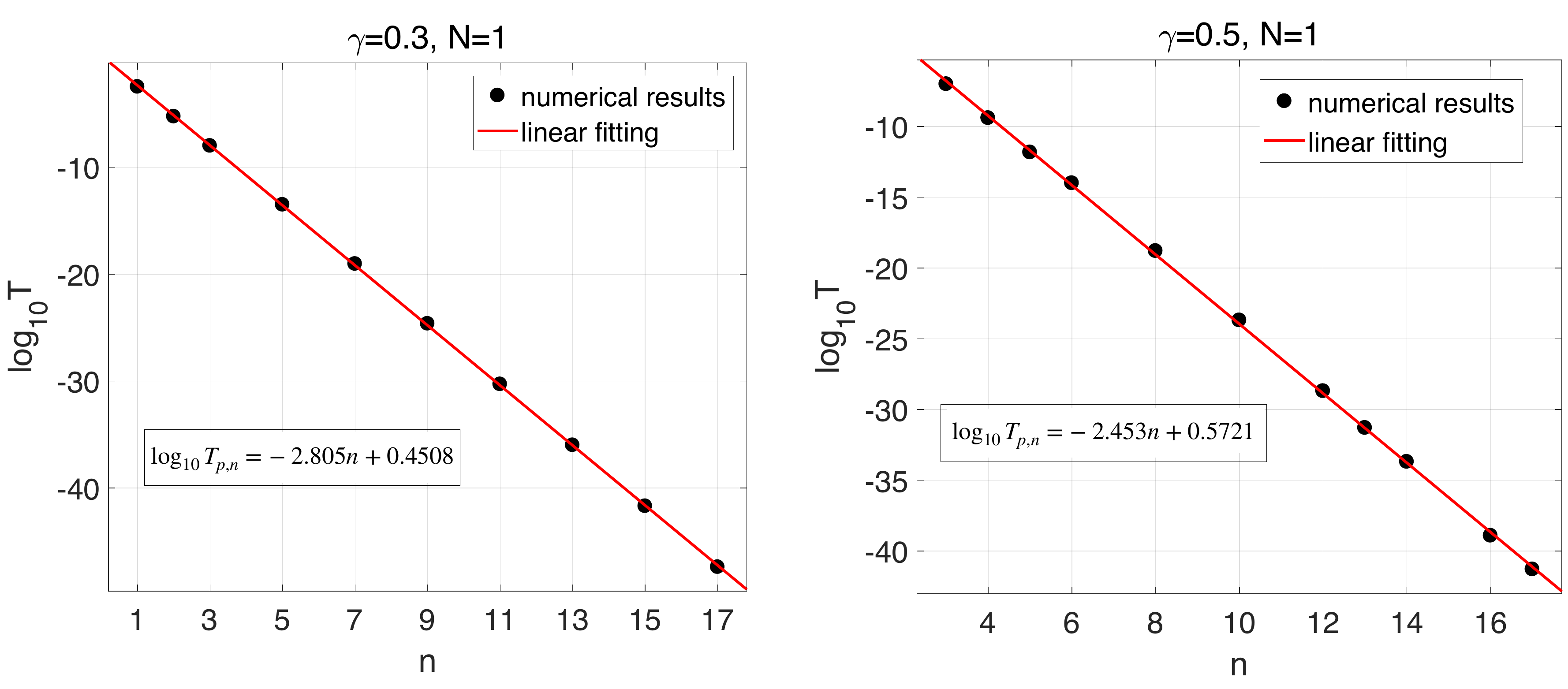}
		\caption{$T_{p,n}$ as  functions of $n$ for $N=1$ and $\gamma=0.3$ and $0.5$. The black dots are the data, obtained by varying temperature to find the eigenvalue $N=1$, and the red lines are linear fitting of $\log{T_{p,n}}$, with fitting parameters given in the boxes. We clearly see that $T_{p,n}$ scales as  $e^{-A n}$, as we obtained analytically.}
\label{fig_T_1}
	\end{center}
\end{figure*}

In Fig.  \ref{fig_T_1} we plot $T_{p,n}$ with $n$ up to $17$ for particular $N=1$.  We clearly see that $T_{p,n}$ scale as $e^{-A n}$, as in Eq. (\ref{ch_18}).  We extracted $\beta_{N=1}$ from the  fit to the exponential form and obtained $\beta_{N=1} = 1.62$ for $\gamma =0.3$ and $\beta_{N=1} = 1.12$ for $\gamma =0.5$. These values are quite close to the exact values, extracted from Eq. (\ref{su_15_2}): $\beta_{N=1} = 1.71$ for $\gamma =0.3$ and $\beta_{N=1} = 1.27$ for $\gamma =0.5$.
  The small difference comes from the numerical error of the hybrid frequency scale, which effectively shifts
  $N$ up by roughly  $0.07$ for $\gamma =0.3$ and $0.13$ for $\gamma =0.5$.

 \begin{figure*}
	\begin{center}
     \includegraphics[width=12cm]{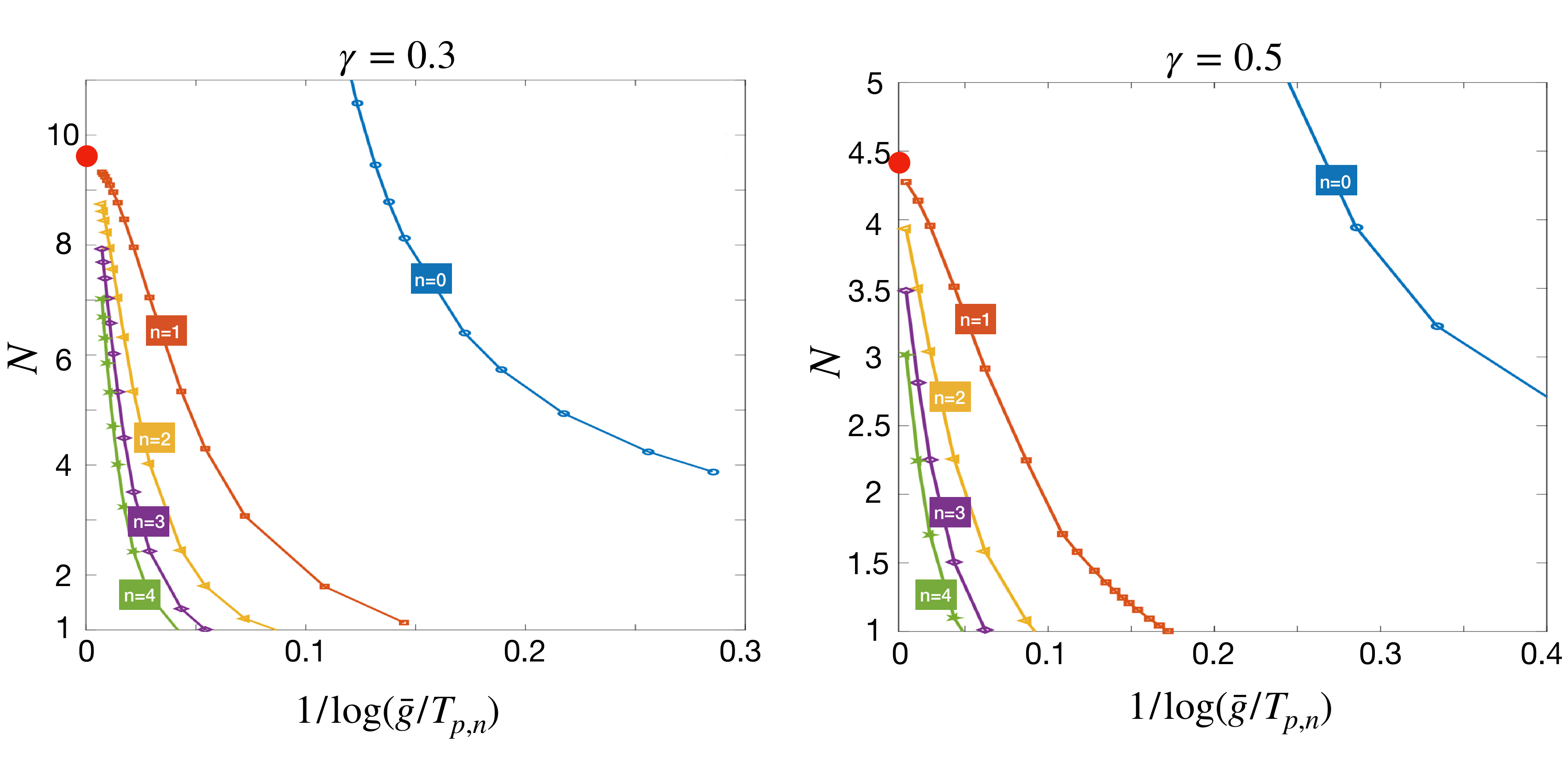}
\caption{The numerical solution of the gap equation  for small but finite $T$ for $\gamma =0.3$ and $0.5$.   The temperatures
 $T_{p,n}$ are the onset temperatures for the pairing in different topological sectors (the corresponding eigenfunctions change sign $n$ times as functions of discrete Matsubara frequency).  The highest $T_{p,0} \propto 1/N^{1/\gamma}$ terminates at $N = \infty$. Analytical reasoning shows that all other $T_{p,n}$  vanish at $N = N_{cr}$ (big red dot). Numerical results show that $T_{p,n}$ with finite $n >0$ indeed approach zero
 at $N=N_{cr}$.}
 \label{fig_T}
 \end{center}
 \end{figure*}

\subsection{The structure of the gap function, $\Delta_n (m)$}
\label{sec:gap}
\begin{figure*}
	\begin{center}
		\includegraphics[width=10cm]{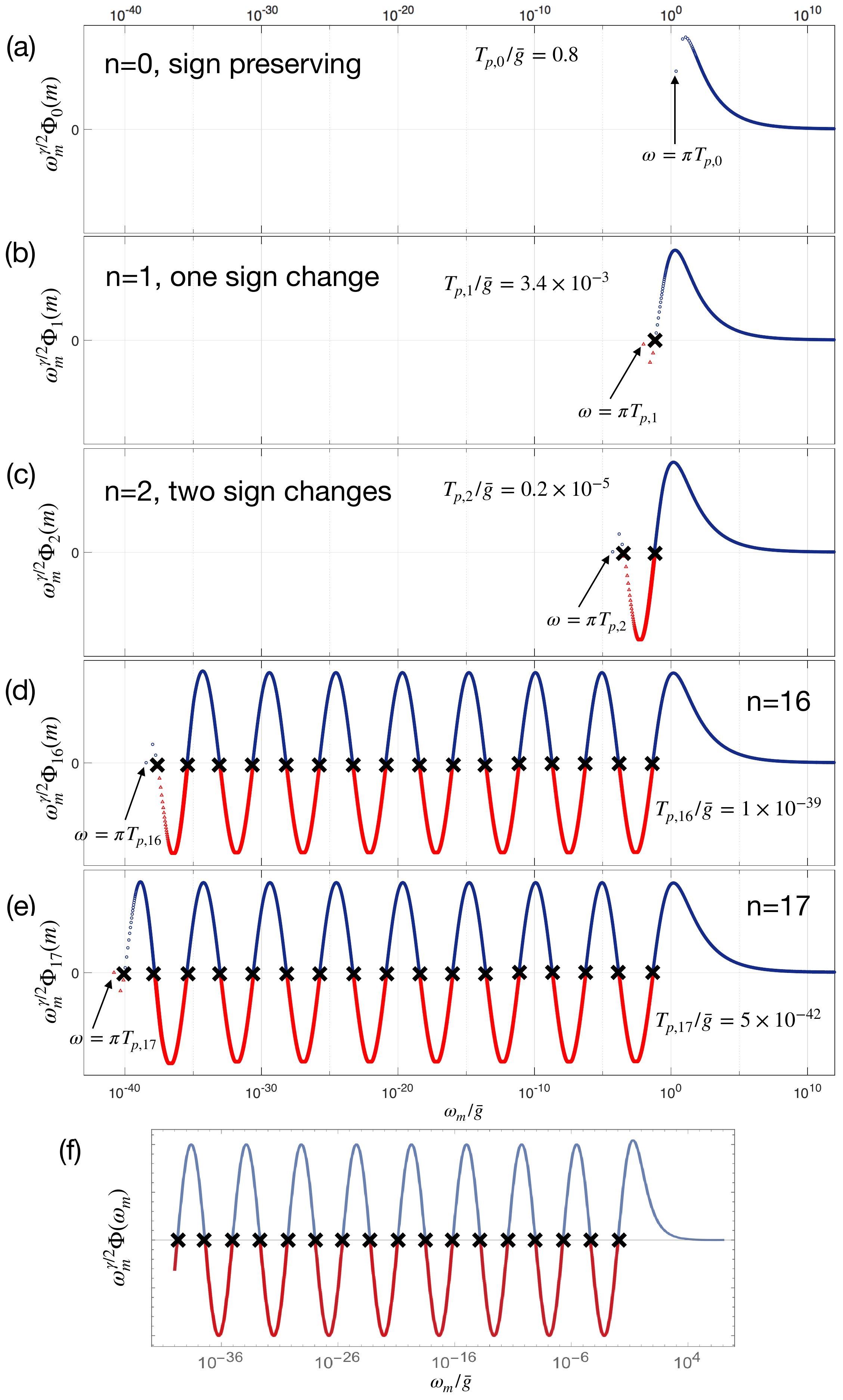}
		\caption{Panels (a)-(e) -- The pairing vertex, $\Phi_n (m)$ at $T=T_{p,n}$, as a function of the Matsubara frequency $\omega_m = \pi T (2m+1)$ for representative parameters $\gamma=0.5$ and $N=1$.
We show $\Phi_n (m)$ for $n=0,1,2,16,17$. The corresponding $T_{p,n}$ are shown in the figures.
    For $n=0$,  $\Phi_0 (m) \propto (1|m|^\gamma + 1/|m+1|^\gamma)$ does not change sign.
      Other $\Phi_n (m)$ change sign $n$ times, and  $T_{p,n} \propto e^{-An}$. The results for $n=16,17$ show that $\Phi_n (m)$ oscillates at $m \gg 1$ as a function of $\log{m}$, with the amplitude proportional to $1/|m|^{\gamma/2}$.  Panel(f) - $\Phi (\omega_m)$ from the exact solution of the linearized equation for the pairing vertex at $T=0$. The positions of zeros of $\Phi_n (m)$ are marked by crosses. The smallest frequencies $\omega_0 = \pi T_{p,n}$ (different for different $n$) are shown by arrows. }\label{fig_T_2}
	\end{center}
\end{figure*}
\begin{figure*}
	\begin{center}
		\includegraphics[width=12cm]{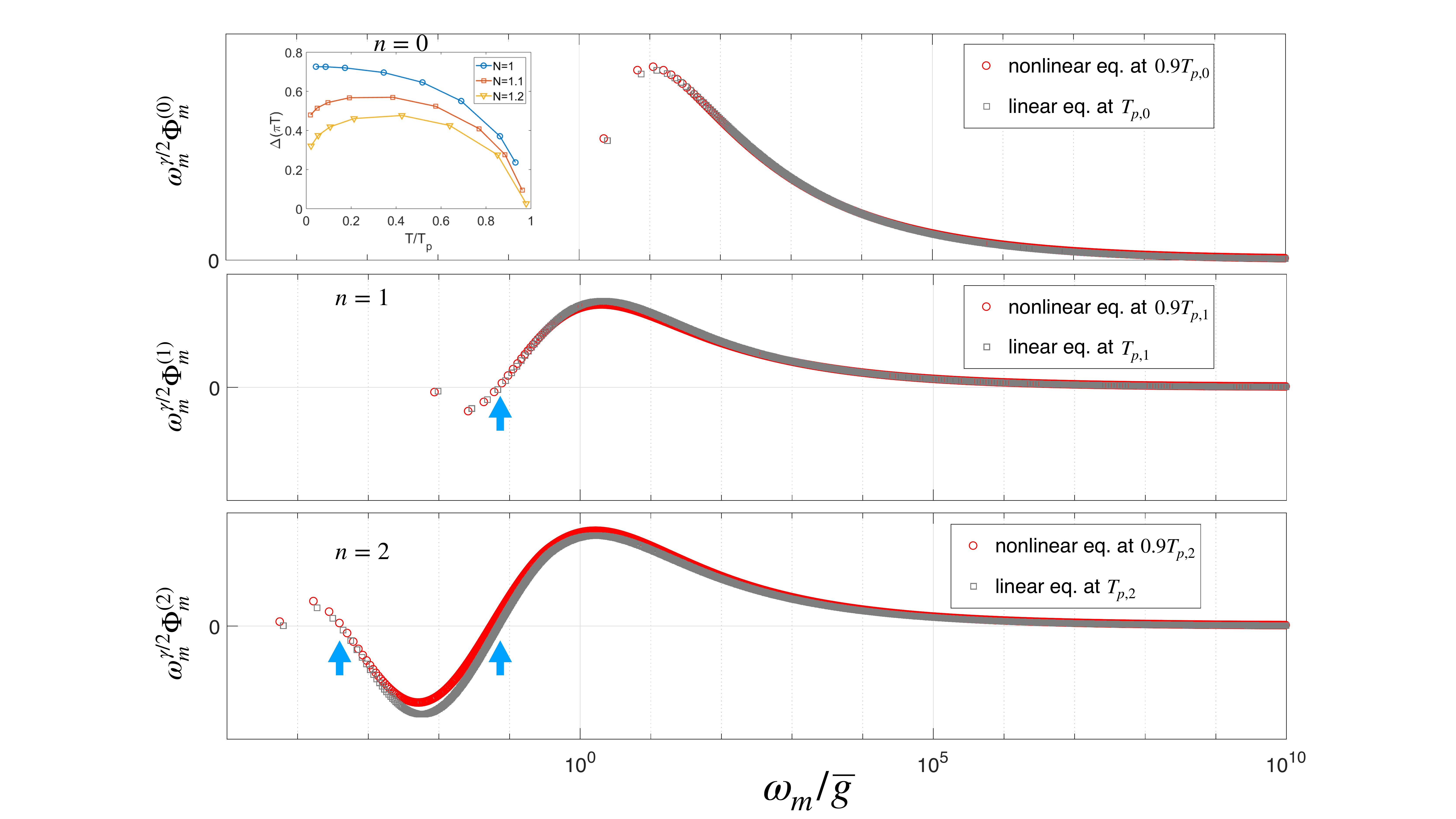}
		\caption{The solution of the nonlinear equation for the pairing vertex for $T = 0.9 T_{p,n}$, along with the solution at $T=T_{p,n}-0$. The three panels are the solutions for $n=0,1,2$.  The number
 of sign changes remains the same at $T_{p,n}$ and $0.9 T_{p,n}$, as indicated by the blue arrows, and the frequencies, at which $\Phi_n (m)$  changes sign,  do not shift with $T$.}
\label{fig_T_3}
	\end{center}
\end{figure*}

Another result of the $T=0$ analysis is that the solutions of the nonlinear gap equation with different $n$ are topologically distinct --- the gap function $\Delta_n (\omega_m)$ changes sign $n$ times as a function of Matsubara frequency.  Because we expect that $\Delta_n (m)$, which develops below $T_{p,n}$, becomes
$\Delta_n (\omega_m)$ at $T=0$, it should change sign $n$ times as a function of Matsubara number $m$. The same should be true for the pairing vertex $\Phi_n (m)$.  In Fig.\ref{fig_T_2} we show $\Phi_n (m)$ for a few smallest $n$ and for $n=16,17$. We see that $\Phi_n(m)$ indeed changes sign $n$ times.  At large $n$, $\Phi_n (m)$ oscillates at large $m$  as a function of $\log{m}$, with the amplitude proportional to $1/|m|^{\gamma/2}$. This is exactly the same behavior as in
 Eq. (\ref{ch_15_c}), given that $\Delta_n (m) \sim \Phi_n (m) |m|^\gamma$.  For comparison, in the last panel of Fig. \ref{fig_T_2} we plot the exact $\Phi (\omega_m)$ at $T=0$. We see that at $T= T_{p,n}$, the form of $\Phi_n (m)$ for $m \gg 1$  is quite similar to that at $T=0$.

 That $\Phi_n (m)$ has to change sign at least once  follows from the relation between $\Phi (0)$ and $\Phi (m>0)$,
 Eq. (\ref{ch_1}). For  $K \gg N$:
  \beq
  \Phi (0)  \approx - \frac{1}{K} \sum_{m=1}^\infty \frac{\Phi (m)}{A (m) + \frac{2m+1}{K}} \left(\frac{1}{m^\gamma} + \frac{1}{(m+1)^\gamma}\right)
  \label{ch_1_a}
  \eeq
 This relation shows that even if $\Phi (m)$ has the same sign for  all $m>0$, $\Phi (0)$  would still be of opposite sign. This is consistent with Fig. \ref{fig_T_2}, which shows that $\Phi_1 (m)$ changes sign at
$m=O(1)$
 and  keeps the same sign at larger $m$. The same holds for larger $n$ - the first sign change occurs at $m = O(1)$, i.e., at $\omega_m \sim T_{p,n}$.
   In Fig. \ref{fig_T_3} we show the results for $\Phi_n (m)$ for $n=0,1,2$ obtained by solving the nonlinear  equation for the pairing vertex for $T \leq T_{p,n}$. We expanded to order $\Phi^3_m$ and
  used the solution at $T= T_{p,n}$ as the source. We see that the number of sign changes remains the same,
   and the frequencies, at which the sign of $\Phi_n (m)$ changes,  remain essentially independent on $T$.
    This is consistent with the result in paper I that at $T=0$,  $\Delta_n (\omega_m)$ changes sign $n$ times at finite $\omega_m$.

\begin{figure}
	\includegraphics[width=\columnwidth]{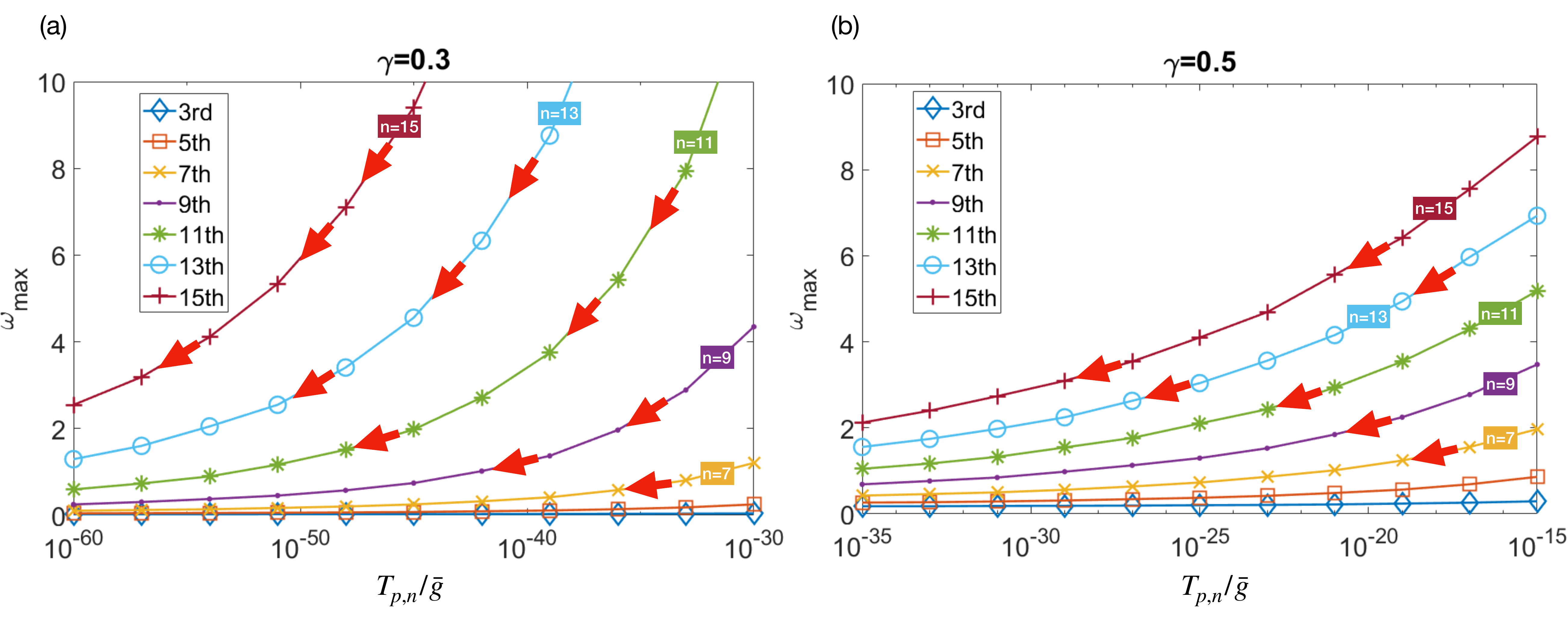}
	\caption{The highest frequency $\omega_{max}$, at which $\Delta_n (m)$ changes sign, plotted vs $T_{p,n}$ for various $n$. The arrows indicate the direction towards smaller $T_{p,n}$. We see that $\omega_{max}$ decreases together with $T_{p,n}$, i.e., $n$ sign changes of $\Delta_n (m)$ occur at progressively smaller Matsubara frequencies.}\label{fig:omegamax}
\end{figure}

Finally, in Fig.\ref{fig:omegamax} we follow $\Delta_n (m)$ along the line $T_{p,n} (N)$ and show the evolution of the
  frequency $\omega_{max}$, at which $\Delta_n (m)$  changes sign for the last time (i.e., all sign changes are at $\omega_m < \omega_{max}$).  Because all $T_{p,n}$ terminate at $N = N_{cr}$,
     $\omega_{max}$ must shrink, as $T_{p,n}$ decreases, and vanish at $T_{p,n} \to 0$, because right at $N = N_{cr}$ and $T=0$,  $\Delta (\omega_m)$ is sign-preserving (see paper I).   The data in Fig.\ref{fig:omegamax} show that $\omega_{max}$ indeed decreases with decreasing $T_{p,n}$ for all values of $n$, as long as $n$ remains finite.

\section{Away from a QCP}
\label{sec:away_from_a_qcp}

 We now analyze how $T_{p,n}$ and the DoE at $T \to 0$ change  away from a QCP, when the pairing boson acquires a finite mass, $M_b$.  We argue that a finite $M_b$ introduces qualitative changes in the system behavior, i.e.,
  there is a qualitative difference between the structure of the DoE at a finite $M_b$ and at  $M_b =0$.  Specifically, we argue that a finite $M_b$ (i) makes the number of $T_{p,n}$ for a given $N$ finite and (ii) splits $T_{p,n}$ at the smallest $T$ such that different $T_{p,n}$ terminate at different $N_{cr,n}$, all smaller than $N_{cr}$.   The temperature $T_{p,0}$ is still nonzero for any $N$, but  at a finite $M_b$ it acquires a conventional form
  $T_{p,0} \sim M_b e^{-1/\lambda}$, where $\lambda \propto 1/N$ [see Eq. (\ref{chh_3}) below].

   We begin with the analytical analysis.  We model the bosonic propagator away from a QCP by
  \beq
  V
   (\Omega_m) = \frac{{\bar g}^\gamma}{\left(\Omega^2_m + M^2_b\right)^{\gamma/2}}
  \label{chh_1}
  \eeq
  The linearized gap equation becomes
    \beq
   \Delta (\omega_m) = \frac{{\bar g}^\gamma}{N} \pi T \sum_{m' \neq m} \frac{\Delta (\omega'_{m}) - N \Delta (\omega_m) \frac{\omega'_{m}}{\omega_m}}{|\omega'_{m}|} ~\frac{1}{\left(|\omega_m - \omega'_{m}|^2 + M^2_b\right)^{\gamma/2}}.
     \label{ss_11_b_1}
  \eeq
   Consider first the limit $T \to 0$. Replacing the  summation over frequency by integration with $T$ as the lower limit, we immediately find that at a finite $M_b$ there
   is a simple difference between sign-preserving and sign-changing solutions for $\Delta (\omega_m)$.
   For the sign-preserving solution, $\Delta (0)$ is finite. Taking properly the limit $\omega_m \to 0$,  we obtain
   $T_{p,0}$ from the self-consistent equation on $\Delta (0)$:
   \beq
   \left(1 + \left(\frac{{\bar g}}{M_b}\right)^\gamma\right) = \frac{1}{N} \left(\frac{{\bar g}}{M_b}\right)^\gamma \log{\frac{M_b}{T_{p,0}}}.
  \label{chh_2}
  \eeq
   The second term in the lhs is the contribution from the self-energy, which away from a QCP has a Fermi-liquid form at frequencies below $M_b$.  Solving for $T_{p,0}$, we find
 \beq
   T_{p,0} \sim M_b e^{-N (1+ \lambda)/\lambda},  ~~~  \lambda = 1 + \left(\frac{{\bar g}}{M_b}\right)^{\gamma}
  \label{chh_3}
  \eeq
  We see that $T_{p,0}$ is still nonzero for any $N$, however its dependence on $N$ is exponential. This is similar to the case of a BCS superconductor, where $T_c$ is finite for arbitrary weak coupling, albeit exponentially small. For $N=1$, Eq. (\ref{chh_3}) has the same structure as McMillan formula $T_c \sim \omega_D e^{- (1+ \lambda)/\lambda}$ (Ref. \cite{McMillan}).
   In qualitative distinction to the behavior at a QCP, now the existence of a nonzero $T_{p,0}$ for arbitrary large $N$ is due to ordinary Cooper logarithm in a Fermi-liquid rather than to special properties of fermions with $\omega_m = \pm \pi T$ in a NFL regime. As a consequence, $\Delta_0 (\omega_m)$, emerging below $T_{p,0}$, does not vanish at $T=0$, i.e, at any $N$ the ground state is a superconductor. In this respect, there is no critical $N_{cr}$, separating normal and superconducting states at $T=0$.
 \begin{figure}
	\begin{center}
		\includegraphics[width=12cm]{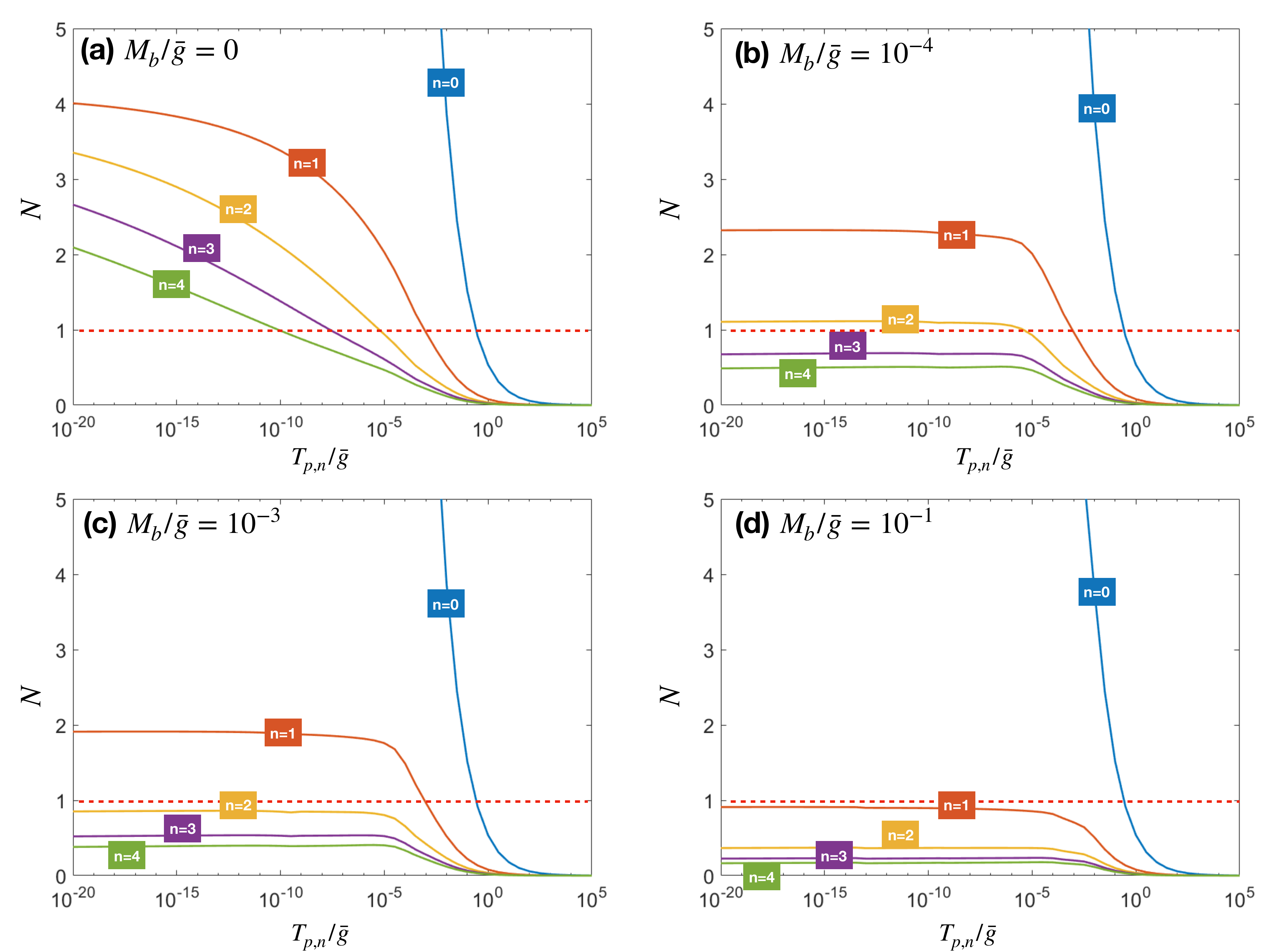}
		\caption{The solutions of the linearized gap equation for a finite boson mass $M_b$.
Different panels are for different $M_b/{\bar g}$, shown in the figures.  We set $\gamma=0.5$.
The critical temperatures $T_{p,n}$ now terminate at different $N_{cr,n}$.  This is qualitatively different from  the behavior at a QCP, where all $T_{p,n}$ with $n >0$ terminate at the same $N = N_{cr}$.}\label{fig:finitemass}
	\end{center}
\end{figure}

 \begin{figure}
	\begin{center}
		\includegraphics[width=10cm]{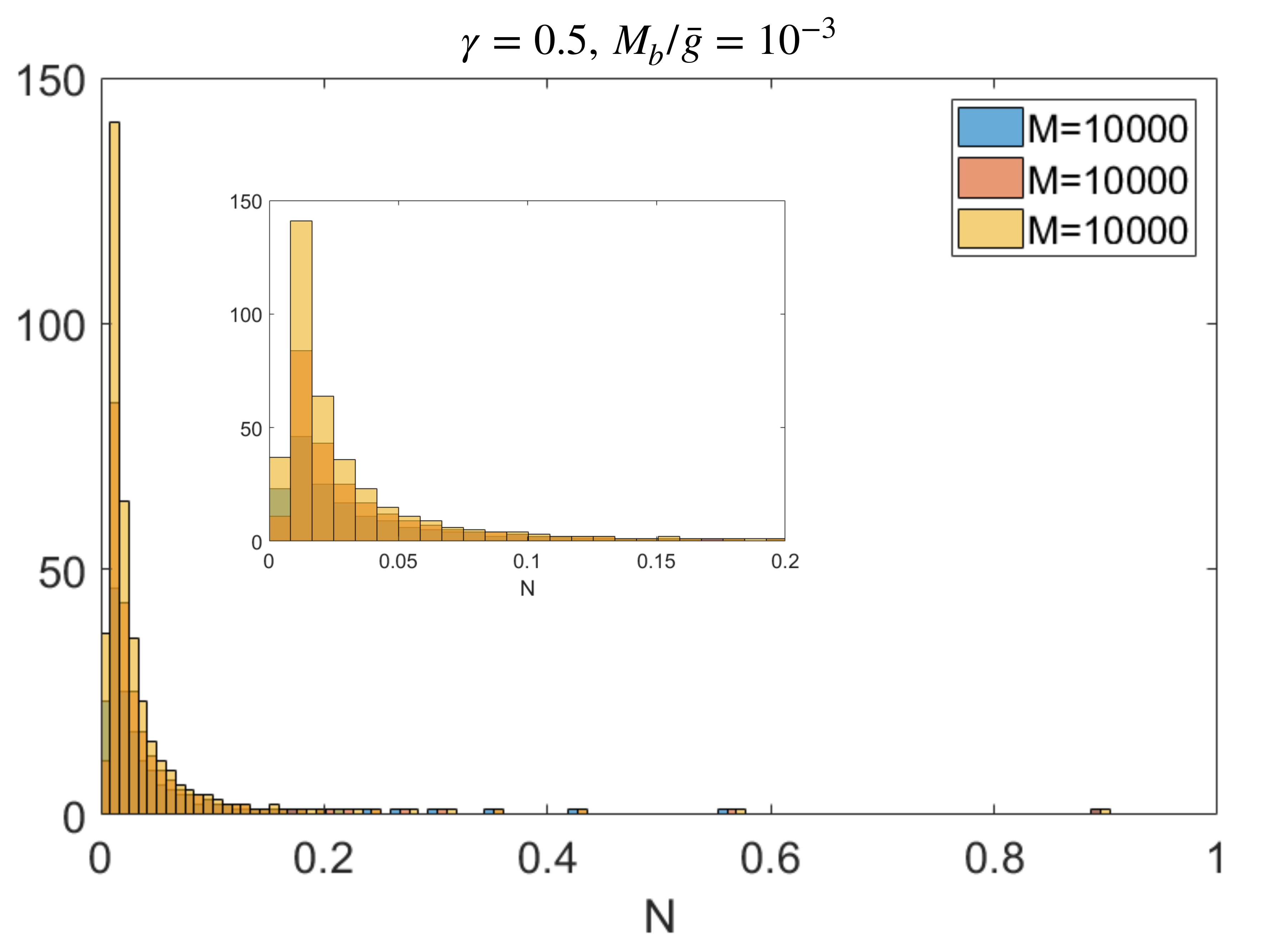}
		\caption{The histogram of the DoE for a finite boson mass $M_b/{\bar g} = 10^{-3}$ and $\gamma=0.5$.
 We see that the histogram is heavily shifted towards $N=0$ because now there is a finite number of points in a  given interval around a particular $N$ even when the total number of sampling Matsubara points $M$ tends to infinity.  This is qualitatively different from the case $M_b =0$ in Fig. \ref{fig:DOShistogram}, where the number of points in a given interval around any $N < N_{cr}$ scales with $M$.}
 \label{fig:histogram_massive}
	\end{center}
\end{figure}
 For solutions with $n >0$, $\Delta_n (\omega_m)$  must vanish at $\omega =0$ because there is just
  a single solution with a finite $\Delta (0)$.
    This  sets the condition
  \bea
  &&\Delta (0) \left(1 + \left(\frac{{\bar g}}{M_b}\right)^\gamma\right) =
\frac{{\bar g}^\gamma}{N} \int_0^\infty \frac{\Delta_n (\omega'_m)}{\omega'_m} \frac{d\omega '_{m}}{\left((\omega'_m)^2 + M^2_b\right)^{\gamma/2}} =0
  \label{chh_4}
  \eea
  We show below that $\Delta_n (\omega'_m)$ scale as $(\omega'_m)^2$ at small $\omega'_m$, hence
    the integral in (\ref{chh_4}) is infra-red convergent.  Eq. (\ref{chh_4}) then
    implies that $\Delta_n (\omega_m)$ must change sign $n$ times at some finite $\omega'_{m}$.
    This is  qualitatively different from the situation at a QCP.  There, all $T_{p,n}$ with finite $n$ terminate  at $T=0$ at $N = N_{cr}$.  The gap function at $N = N_{cr}$   vanishes at $\omega_m =0$, yet it  remains sign-preserving (see paper I for details).   This holds at a QCP  because $\Delta (\omega'_m) \propto |\omega'_m|^{\gamma/2}$, in which case $\Delta (\omega'_m)/\omega'_m$ is singular at $\omega_m \to 0$, and one cannot just set $\omega_m =0$ in the gap equation, as it is done in Eq. (\ref{chh_4}).
  At small  $M_b$,  $T_{p,n}$ approaches zero at $N$ only slightly below $N_{cr}$, and $\Delta_n (\omega'_m)$
    must  recover the gap function  at a QCP at $N=N_{cr}$ at frequencies above some scale, which vanishes when $M_b \to 0$. Because $\Delta (\omega_m)$ at $N=N_{cr}$ is sign-preserving,  the $n$ sign changes of
  $\Delta_n (\omega_m)$ have to occur below this scale.

We now expand in $\omega_m$ the rhs of the gap equation  (\ref{ss_11_b_1}) for $\Delta_{n>0} (\omega_m)$.   Expanding  and using (\ref{chh_4}) to cancel out the leading term, we obtain $\Delta_n (\omega_m) = A_n (\omega_m/M_b)^2$, where $A_n$ is given by
\beq
A_n = \left(\frac{{\bar g}}{M_b}\right)^\gamma \frac{\gamma}{N} \int_0^\infty dx \frac{\Delta_n (x)}{x} \frac{x^2 (1+\gamma) -1}{(x^2+1)^{2+\gamma/2}}
\label{chh_5}
\eeq
where $x = \omega_m/M_b$.   The integral in (\ref{chh_5}) converges at $x = O(1)$, hence by order of magnitude
$A_n  \sim {\bar g}^\gamma/(M^{2+\gamma}_b N)$ with $n-$dependent prefactor.  For an estimate, we assume that at large $n$  the integral is determined by $x_n \sim 1/n$, before oscillations begin, and that $\Delta_n (x <x_n)
 \approx  A_n x^2$.  Substituting into  (\ref{chh_5}), we obtain that the solution is possible only for a given $N \sim 1/n^2$
\beq
 N \gamma = - \frac{x^2_n}{2} \left(\frac{{\bar g}}{M_b}\right)^\gamma
 \int_0^{x_n} dx x \frac{x^2 (1+\gamma) -1}{(x^2+1)^{2+\gamma/2}}
\label{chh_6}
\eeq
The outcome of this analysis is that at a nonzero $M_b$ the solutions  of the linearized gap equation for different $n$ exist at different $N$, i.e., each $T_{p,n}$ terminates at its own $N_{cr,n}$.

In Fig. \ref{fig:finitemass} we show the results of the numerical solution of the gap equation for a finite $M_b$.
 We see that $T_{p,n}$ indeed terminate at different $N$.
 %, which can be both positive and negative.
 We verified that $T_{p,0}$  is exponential in $N$, like in Eq. (\ref{chh_3}).  The  number of solutions, for which $T_{p,n}$ crosses $N=1$, is finite for any nonzero $M_b$. It decreases one by one
  with increasing $M_b$ and vanishes once $M_b$ exceeds some critical value.  At larger $M_b$, there is only one onset temperature $T_{p,0}$ for the physical case $N=1$, and the behavior below $T_{p,0}$ is qualitatively the same as in BCS theory.

The same behavior  shows up in the analysis of the DoE at $T\to 0$. Because there is only a finite number of termination points of $T_{p,n}$ in any finite interval of $N$, the normalized DoE $\nu (N)$ vanishes for all $N \neq 0$ in the formal limit $M \to \infty$, where $M$ is the number of Matsubara points, probed in a numerical calculation.  Because termination lines cluster around $N=0$, and the total $\int dN \nu (N) =1$, the DoE becomes $\nu (N) = \delta (N)$ in this limit.
 In practice, this implies that the DoE will decrease for any finite $N$ when $M$ increases, and the histogram of $\nu (N)$ will shift with increasing $M$ towards $N=0$. In Fig... we show the  numerical results for $\nu (N)$. We see precisely this behavior. Note also that the largest $N_{cr,1}$  shifts down from $N_{cr}$ as $M_b$ increases.

\section{Conclusions}
 \label{sec:conclusions}

In this paper, the second in a series, we continued our analysis of  the interplay between pairing and  NFL in the model of fermions with singular dynamical interaction $V
%_{\rm eff}
 (\Omega_m) \propto 1/|\Omega_m|^\gamma$
(the $\gamma$ model).  We introduced a knob (the parameter $N$) to vary  the relative strength of the dynamical interaction in the particle-hole and particle-particle channels.  In paper I we studied the $\gamma$ model at $T=0$ and $0<\gamma <1$.   We found that superconducting order exists for $N < N_{cr} (\gamma)$ and that for any  such $N$ there is an infinite number of solutions of the nonlinear gap equation, $\Delta_n (\omega_m)$, specified by the number of times ($n$) the gap function changes sign as a function of $\omega_m$.  At large $n$, $\Delta_n (0) = e^{-an}$  is exponentially small in $n$.

   In this paper, we analyzed the same model at a finite $T$. We showed
   the DoE remains continuous at $T \to 0$ and coincides with the one at $T=0$, i.e., the existence of an infinite set of solutions of the gap equation is not a special property of $T=0$.  We further showed that each $\Delta_n (\omega_m)$  vanishes at a critical temperature $T_{p,n} \sim \Delta_n (0)$.  Viewed as functions of $N$, all $T_{p,n} (N)$ with $n >0$ terminate at the same $N = N_{cr}$, which turns out to be a multicritical point (see Fig.\ref{fig:illustration}(b) and Fig.\ref{fig_T}).   The eigen-function $\Delta_n (m)$ at $T = T_{p,n} -0$ changes sign $n$ times as a function of Matsubara number $m$, and retains this property down to $T=0$, where it becomes a continuous function $\Delta_n (\omega_m)$.  The temperature $T_{p,0}$ behaves differently -- it remains finite for all $N$ and scales at large $N$ as $T_{p,0} \sim 1/N^{1/\gamma}$.   In this limit, the pairing at $T_{p,0}$ predominantly involves fermions with the two lowest Matsubara frequencies $\pm \pi T$, because for these fermions nonthermal self-energy vanishes, which implies that pairing does not compete with NFL.
     At $N > N_{cr}$, $\Delta_0 (m)$ initially grows below $T_{p,0}$, but then changes trend and vanishes at $T \to 0$.

This behavior holds only at a QCP.  Away from a QCP, $T_{p,n}$ with different $n$ terminate at different $N_{cr,n}$, and for any $N >0$, the number of $T_{p,n}$ is finite. It drops one by one as the bosonic mass $M_b$ gets larger, and disappears above a certain $M_b$.  The onset temperature  $T_{p,0}$ is still finite for any $N$, but the gap $\Delta_0 (m)$ does not vanish at $T \to 0$, i.e., the ground state is a superconductor  for any value of  $N$.

The next paper will be devoted to analysis of how the physics of the $\gamma$ model at a QCP changes between  $\gamma < 1$ and $\gamma >1$.

 \acknowledgements
  We thank   I. Aleiner, B. Altshuler, E. Berg, D. Chowdhury, L. Classen, R. Combescot, K. Efetov,  R. Fernandes,  A. Finkelstein, E. Fradkin, A. Georges, S. Hartnol, S. Karchu, S. Kivelson, I. Klebanov, A. Klein, R. Laughlin, S-S. Lee, G. Lonzarich, D. Maslov, F. Marsiglio, M. Metlitski, W. Metzner, A. Millis, D. Mozyrsky, C. Pepin, V. Pokrovsky,  N. Prokofiev,  S. Raghu,  S. Sachdev,  T. Senthil, D. Scalapino, Y. Schattner, J. Schmalian, D. Son, G. Tarnopolsky, A-M Tremblay, A. Tsvelik,  G. Torroba,  E. Yuzbashyan,  and J. Zaanen for useful discussions.   The work by  AVC was supported by the NSF DMR-1834856.

 \appendix

\section{Free energy below $T_{p,0}$ at large $N$}
\label{appendix_1}

In this Appendix we address one issue about the system behavior at large $N$, below $T_{p,0}$.   As we said in the text and in earlier papers~ \cite{Wu_19_1,Abanov_19,Chubukov_2020a},
$T_{p,0}$
remains finite  for arbitrary large $N$ because the pairing in the large $N$ limit predominantly involves fermions with Matsubara frequencies $\omega_m = \pm \pi T$ (Matsubara numbers $m =0,-1$), for which in Eliashberg theory the nonthermal part of the self-energy vanishes. Because only the nonthermal part of the self-energy appears in the gap equation, these fermions can be viewed as free quasiparticles for the purpose of the pairing. For all other fermions, nonthermal self-energy is strong and acts against pairing.  We argue that the pairing at $T_{p,0}$ involves fermions with $\omega_m = \pm \pi T$, which develop $\Delta_0 (m=0) = \Delta_0 (-1)$ below $T_{p,0}$. A much smaller gap $\Delta_0 (m)$ at other Matsubara frequencies is then induced by proximity.  The gap $\Delta_0 (0)$ initially increases as $T$ decreases, but never exceeds $T$. At the smallest $T$,  $\Delta_0 (0)$ decreases proportional to $T$ and vanishes at $T=0$, where a Matsubara frequency becomes a continuous function, and the special role of frequencies $\pm \pi T$ is lost.

The very existence of the solution of the Eliashberg equations with a nonzero $\Delta_0 (m)$ implies that the free energy of such state, $F_{p,0}$, is smaller than that of the normal state (see Paper I and
 Refs. \cite{eliashberg,wada,*maki4,*Bardeen,*haslinger,with_emil}).
 In explicit form, in this case
\beq
\delta F = F_{p,0}-F_n=-\frac{{\bar g}^2\nu_0}{2N^{1+2/\gamma}}t^{2-\gamma}(t^\gamma-1)^2
\eeq
 where $t= T/T_{p,0}$ and $\nu_0$ is the density of states at the Fermi level in the normal state.

\begin{figure}
	\begin{center}
		\includegraphics[width=12cm]{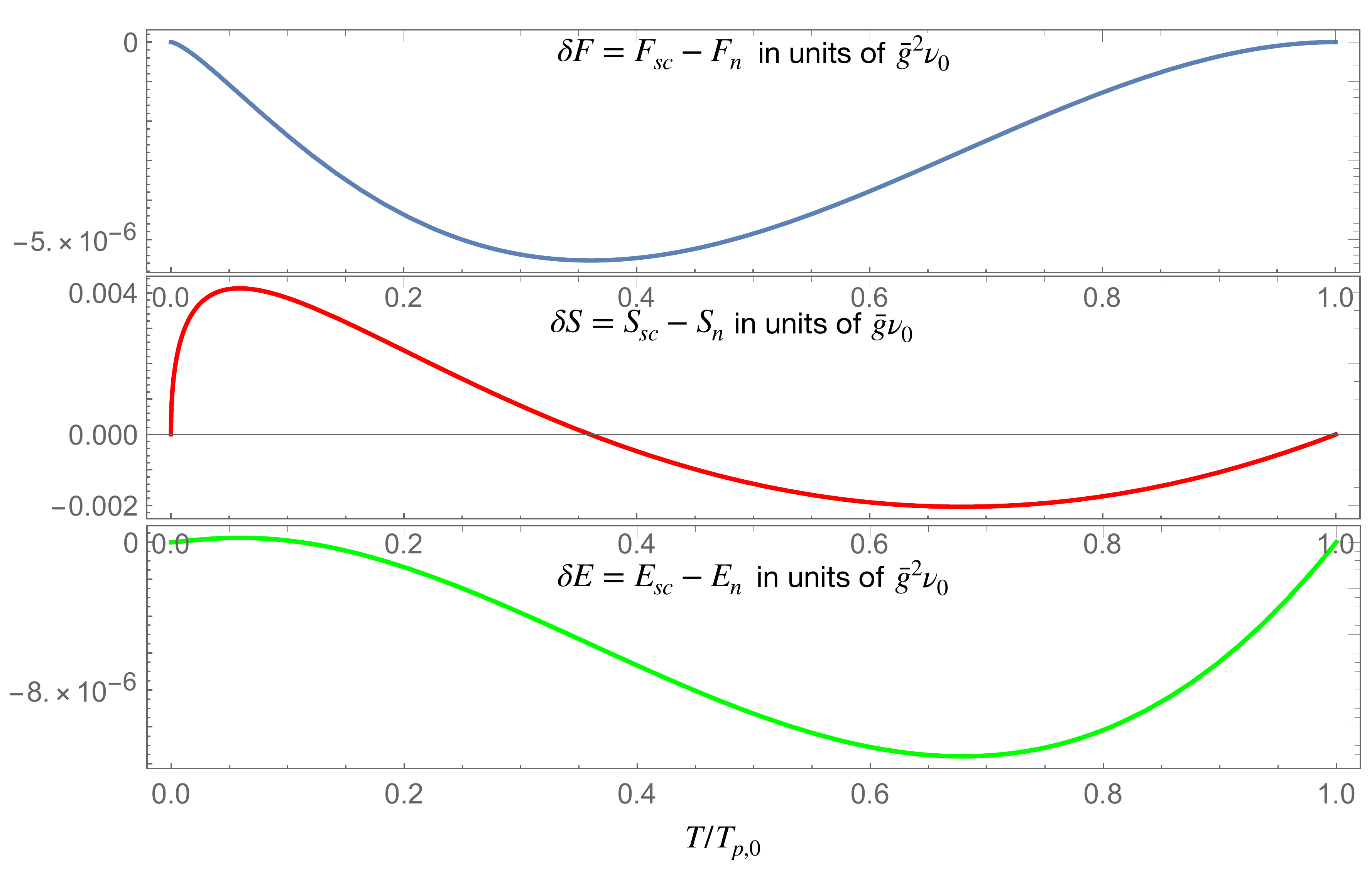}
		\caption{The difference of free energy, entropy and energy between the SC state and normal state, as a function of temperature. We choose $\gamma=0.5$ and $N=5>N_{cr}$.}\label{fig:Free}
	\end{center}
\end{figure}

  The issue we address here is whether a negative  $\delta F = \delta E - T \delta S$ comes from the change of the energy, like in BCS superconductor, or from the change of the entropy.
  In our case, we can compute both terms explicitly at large $N$.  We obtain
\begin{equation}
  \begin{aligned}
  \delta S&=S_{p,0}-S_n=\frac{\bar g
  \nu_0\pi}{N^{1+1/\gamma}}t^{1-\gamma}(t^\gamma-1)\left(t^\gamma(\gamma+2)+\gamma-2\right)\\
		\delta E&=E_{p,0}-E_n=\frac{{\bar g}^2\nu_0}{2N^{1+2/\gamma}}t^{2-\gamma}(t^\gamma-1)(t^\gamma(\gamma+1)+\gamma-1)
	\end{aligned}
\end{equation}
We plot $\delta F, \delta S$, and $\delta E$ in Fig.\ref{fig:Free} for a particular case of $\gamma=0.5$ and $N=5$ $N_{cr} (\gamma =0.5) = 4.476 <5$.
 We clearly see that all three quantities vanish at $T=0$, as expected.   We also see that there is a clear change of behavior at  larger $t$, when $\Delta_0 (m)$ increases with decreasing $t$, and at smaller $t$, where $\Delta_0 (m)$ scales with $t$.  In the first case, $\delta E$ is negative and $\delta S$ is positive, i.e.,
   negative $\delta F$ is due to negative $\delta E$, as in a BCS superconductor.  However, at smaller $t$,
   $\delta E$ is reduced and actually becomes positive at the smallest $t$.  In this $t$ range, a negative $\delta F$ is entirely due to positive $T\delta S$, i.e smaller free energy for the state with a finite $\Delta_0 (m)$ is an
   entropic effect. This is also consistent with the fact that the gap function vanishes at $T=0$.

\section{Onset temperatures $T_{p,n}$ at small $\gamma$}
\label{appendix_2}

The limit $\gamma \to 0$ of the $\gamma$ model attracted a lot of attention from various sub-communities in physics\cite{raghu_15,Wu_19,Wang_H_17,Wang_H_18,Fitzpatrick_15,max_last,senthil,son,*son2} and has been  analyzed in both Eliashberg-type and renormalization group approaches.
 In this Appendix, we present two expressions for $T_{p,n}$ at small $\gamma$ and $N = O(1)$, which is much smaller than $N_{cr} =4/\gamma$.   One  expression is for the pure $\gamma$ model with $V(\Omega) = ({\bar g}/|\Omega|)^{\gamma}$. Another is for the case when we
 introduce an upper cutoff for $V(\Omega_m)$ at $\Omega_m = \Lambda$,
  i.e.,
  modify the interaction to
  \beq
  V(\Omega_m) = \left(\frac{{\bar g}}{|\Omega_m|}\right)^{\gamma} \left[1- \left(\frac{|\Omega_m|}{\Lambda}\right)^{\gamma}\right]
  \label{last_1a}
  \eeq
  for $|\Omega_m| < \Lambda$, and $V(\Omega_m) =0$ for $|\Omega_m| > \Lambda$.
 The distinction is important for small $\gamma$ because at $\gamma \to 0$
   $V(\Omega) \propto |\Omega|^{-\gamma}$ becomes frequency independent, and the pairing problem within the $\gamma$ model becomes equivalent to BCS, but without the upper cutoff for the interaction.
   In this situation, the cutoff at $\Lambda$ is necessary to avoid ultra-violet singularities.  At the same time, at a finite $\gamma$, the gap equation is free from ultra-violet singularities already without $\Lambda$, and if $T_{p,n}$ are smaller than $\Lambda$, the cutoff can be safely neglected.

 We consider the cases when $\Lambda$ is set to infinity (the pure $\gamma$ model) and when $\Lambda$ is finite separately, and then obtain an interpolation formula linking the two cases.

 \subsection{$T_{p,n}$ in the pure $\gamma$-model}

 The point of departure for the calculation of $T_{p,n}$ at small $\gamma$ is the observation in Paper I and in earlier works
 ~\cite{raghu_15,Wang2016,max_last,Wang_H_17,Wang_H_18,Wu_19}  that for $N= O(1)$,  $T_{p,0}$  is  determined by fermions with  frequencies  much  larger than ${\bar g}$. For these fermions, the normal state self-energy at $T=0$, $\Sigma (\omega_m) \approx |\omega_m|^{1-\gamma} {\bar g}^\gamma \text{sgn} \omega_m $ is smaller than the bare $\omega_m$ and can be safely neglected.   Without the self-energy, $\Phi (m) = \Delta (m)$ and the gap equation reduces to
  \beq
    \label{eq:gapeq_d_1}
    \Delta (m) = \frac{K}{N} \sum_{m' \neq m} \frac{\Delta (m')}{|2m'+1|}
    ~\frac{1}{|m - m'|^\gamma}.
    \eeq
where, we remind the reader, $K = ({\bar g}/(2\pi T))^\gamma$.
For $m, m' \gg 1$, the summation over $m'$ can be replaced by the integration. Also, for small $\gamma$,
$1/|m-m'|^\gamma$ is reasonably well approximated by $1/|m|^\gamma$ for $|m| > |m|'$ and by $1/|m'|$ for $|m'|>|m|$.  As a result, the gap equation for $m \gg 1$ can be approximately re-expressed as
 \beq
    \label{eq:gapeq_d_2}
    \Delta (m) = \frac{K}{N} \left(~\frac{1}{|m|^\gamma} \int_0^m dm' \frac{\Delta (m')}{m'} + \int_m^\infty dm' \frac{\Delta (m')}{(m')^{1+\gamma}}\right)
 \eeq
Introducing $z = m^\gamma$ and differentiating twice over $z$, we obtain the differential equation
\beq
    \label{eq:diff}
    (z \Delta (z))'' = - \frac{K}{N\gamma} \frac{\Delta (z)}{z^2}
    \eeq
 The solution of this equation must satisfy two boundary conditions. One follows from \ref{eq:gapeq_d_1}:
   $\Delta (z \to \infty) \propto 1/z$.   Another comes from the boundary at $m = O(1)$, i.e., at $z \approx 1$, where the discretization in Eq. \ref{eq:gapeq_d_1} becomes relevant. Sign-preserving $\Delta_0 (m)$ is flat
   at $m = O(1)$ and deviates from $\Delta_0 (0)$ only by $O(\gamma)$. Then, to leading order in $\gamma$,
   the second boundary condition is $\Delta' (z=1) =0$.  For sign-changing $\Delta_n (m)$,
    we can use the fact that the gap function changes sign at $m = O(1)$ and use $\Delta (z=1) =0$.

The generic solution of (\ref{eq:diff}) is
\beq
\Delta (z) =   \frac{1}{\sqrt{z}} \left[A_1  J_1 \left(2\sqrt{\frac{K}{N \gamma z}}\right) + A_2 Y_1 \left(2 \sqrt{\frac{K}{N \gamma z}}\right)\right],
 \label{su_11}
 \eeq
 where $J_1$ and $Y_1$ are Bessel and Neumann functions.

  At large $z$, $J_1 \propto 1/\sqrt{z}$ and $Y_1 \propto \sqrt{z}$.
    To satisfy the boundary condition at $z \to \infty$ we must set $A_2=0$.  Then
 \beq
\Delta (z) =   \frac{A_1}{\sqrt{z}} J_1 \left(2\sqrt{\frac{K}{N \gamma z}}\right).
 \label{su_11_1}
 \eeq
  A similar result has been obtained in Ref. \cite{Wang_H_17}.
Using the second boundary condition for sign-preserving solution ($n=0$ in  our nomenclature) and using
$(x J_1 (x))' = x J_0 (x)$,  we obtain
 \beq
 J_1 \left(2\sqrt{\frac{K}{N \gamma}}\right) =0,
 \eeq
 where, we remind the reader, $K = ({\bar g}/(2\pi T))^\gamma$.
 The highest $T$ at which this equation has the solution is $T = T_{p,0}$, where
 \beq
 T_{p,0} = c {\bar g} (1.4458 N \gamma)^{-1/\gamma}
 \label{nn_1}
 \eeq
  where $c = O(1)$.  For $N=1$ this agrees with
   Ref. \cite{max_last} (see also earlier results~\cite{Wang2016,with_emil}).
  We see that $T_{p,0}$ is indeed much larger than ${\bar g}$ and tends to infinity when $\gamma \to 0$.
   As we said, this is the consequence of the fact that  at $\gamma \to 0$ the $\gamma$ model reduces to the BCS model without the cutoff on the interaction.

   The prefactor $c$ has been computed in Ref. \cite{Wu_19}, and the result is
  $c = 0.502$.  The gap at $T=0$,  $\Delta_0 (\omega_m =0)$ has been obtained in Paper I and in Ref.~\cite{Wu_19}: $\Delta_0 (0) = 0.885 {\bar g} (1.4458 N \gamma)^{-1/\gamma}$. The ratio $2\Delta/T_c = 3.53$, as in  BCS theory.

  For sign-changing solutions, the boundary condition $J_1 (2 \sqrt{K/N\gamma}) =0$ yields
  \beq
 T_{p,n} = c_n {\bar g} (a_n N \gamma)^{-1/\gamma}
 \label{nn_1_1}
 \eeq
  where $c = O(1)$ and $a_1 = 3.6705$, $a_2 = 12.3047$, etc.
  This formula is valid up to some $n$, for which $T_{p,n} = O({\bar g})$. For larger $n$, $T_{p,n} \propto e^{-An}$, like in Eq. (\ref{ch_18}).   The ratio $2 \Delta_n (0)/ T_{p,n}$ is of order 1 for $n = O(1)$, but we did not compute it explicitly.

  Note that although $T_{p,n}$ with $n = O(1)$ are much larger than ${\bar g}$, the ratio $T_{p,1}/T_{p,0} \sim (0.394)^{1/\gamma}$ is parametrically small and vanishes when $\gamma \to 0$.

 \subsection{$T_{p,n}$ in the $\gamma$ model with the infra-red cutoff}

We now consider the opposite limit when $\gamma \to 0$, but the  upper cutoff for $V(\Omega_m)$
 is a finite $\Lambda$. In this case, we use Eq. (\ref{last_1a}) for the interaction.

 The derivation of the differential equation proceeds along the same lines as before, and the result is the same Eq. (\ref{eq:diff}) as before.  However, now
the solution, $\Delta (z)$,
has to satisfy the boundary condition  $\Delta (K^*) =0$, where $K^* = (\Lambda/2\pi T)^\gamma$.
Such a solution is
\bea
&&\Delta (z) \propto \frac{1}{\sqrt{z}} \times  \nonumber \\
&& \left[J_1 \left(2\sqrt{\frac{K}{N \gamma z}}\right)
Y_1 \left(2 \sqrt{\frac{K}{N \gamma K*}}\right)
- Y_1 \left(2\sqrt{\frac{K}{N \gamma z}}\right)
J_1 \left(2 \sqrt{\frac{K}{N \gamma K*}}\right)\right].
\label{last_3a}
\eea
In the limit $K/(N\gamma K^*) = ({\bar g}/\Lambda)^\gamma/(N \gamma) \gg 1$, valid when $\gamma \to 0$ at a finite $\Lambda$,
  we use the asymptotic forms of Bessel and Neumann functions at large values of the argument:
\bea
&& J_1 (x) \approx \sqrt{\frac{2}{\pi x}} \cos{(x -3\pi/4)}, \nonumber \\
&& Y_1 (x) \approx \sqrt{\frac{2}{\pi x}} \sin{(x -3\pi/4)},
\label{last_4a}
\eea
and obtain
\beq
\Delta (z) \propto \frac{1}{z^{1/4}} \sin{\left(2\sqrt{\frac{K}{N \gamma K^*}} - 2 \sqrt{\frac{K}{N \gamma z}}\right)}
\label{last_5a}
\eeq
To the leading order in $\gamma$, this reduces to
\beq
\Delta (m) \propto \sin{\left(\sqrt{\frac{\gamma}{N}} \log{\frac{2 \pi T |m|}{\Lambda}}\right)}
\label{last_6a}
\eeq
Using the second boundary condition for sign-preserving solution, we find
\beq
T_{p,0} \sim \Lambda e^{-\pi/2\sqrt{\lambda_\gamma}},  ~~ \lambda_\gamma = \frac{\gamma}{N}
\label{nn_3}
\eeq
This agrees with Refs. \cite{son,*son2,max_last}.
For $n >0$, using $\Delta (z=1) =0$, we obtain
\beq
T_{p,n} \sim \Lambda e^{-n\pi/\sqrt{\lambda_\gamma}},  ~~n =1,2...
\label{nn_4}
\eeq
Again, $T_{p,1}$ is parametrically smaller than $T_{p,0}$, the ratio $T_{p,1}/T_{p,0}$ vanishes at $\gamma \to 0$ and $T_{p,n}$ with larger $n$ are even smaller.
Note that $T_{p,n}$ is independent on $\gamma$.

The zero-temperature gap $\Delta_n (\omega_m=0)$ in the same limit has been computed in paper I.
 Comparing it with $T_{p,n}$, Eqs. (\ref{nn_3}) and (\ref{nn_4}), we find that $T_{p,n}$ and $\Delta_n (0)$ are of the same order.  A higher accuracy of calculations is required to compute $2\Delta/T_p$ ratios.

\subsection{Intermediate regime}

We now argue that Eq. (\ref{last_3a}),
 describes the gap functions in both, the $\gamma$ model with the cutoff at $\Lambda$, and the pure $\gamma$ model. Indeed, in the opposite limit of large $\Lambda$ and finite $\gamma$,  $K/(N\gamma K^*) = ({\bar g}/\Lambda)^\gamma/(N \gamma) \ll 1$.
 In this case
$ Y_1 \left(2 \sqrt{\frac{K}{N \gamma K*}}\right) \gg J_1 \left(2 \sqrt{\frac{K}{N \gamma K*}}\right)$. Then only $Y_1 \left(2 \sqrt{\frac{K}{N \gamma K*}}\right)$ should be kept in (\ref{last_3a}), and we recover
Eq. (\ref{su_11_1}).   In this limit, the functional form of $\Delta (z)$ does not depend on $\Lambda$ and we recover the result for $T_{p,n}$ for the pure $\gamma$ model.
The crossover between the two regimes occurs at $({\bar g}/\Lambda)^\gamma/(N \gamma) = O(1)$.

\section{The hybrid frequency scale}
\label{app:A}

 Within our computation capacity, the maximum size of frequency mesh
  we can take, is approximately $10^4$.
  %  with this limitation if
   If we
  directly sum over
  Matsubara frequencies
  %, namely
  $\omega_m=(2m+1)\pi T$, the maximum frequency we can
  %touch
    reach is around $10^5T$. Because we are interested in frequencies $\omega_m < {\bar g}$,
     %which means that
      the lowest temperature one can reach is $T\sim 10^{-5}\bar g$.
       %without losing the frequency range in which NFL physics dominates.
 To access exponentially lower temperatures, one can in principle use a logarithmical frequency mesh in which $\log (\omega_m/\pi T) \propto m$, but in this scheme one loses the special role of the first Matsubara frequency
  and, more generally, of Matsubara frequencies with $m = O(1)$, which, e.g., set the phase in the expression for  $\Delta (\omega_m)$ at $T \to 0$, Eq. (\ref{ch_15_b})
  To overcome this difficulty we adopt
 a hybrid frequency scaling.
  Namely, we set \begin{equation}
		\omega_m=
\begin{cases}
(2m+1)\pi T,~m<m_L\\
(2m+1)\pi T+e^{km-b}\pi T, m\geq m_L\\
\end{cases}
\end{equation}
Here $n_L$ is some number, below which we adopt the original  formula for a Matsubara frequency, and beyond which we add an exponentially growing term.  In practice, we have taken
 $m_L\sim 0.1 M$,
  where $M$  is the total number of frequency points.
  One should also properly choose the parameters $k$ and $b$ such that when $m\to m_L$, this exponential term can be neglected compared to $\pi T$ and when $m\to M$,
   the exponential term dominates over the linear term and can reach our upper limit of frequency.
 The change of the frequency form also induces a corresponding change to the measure of summation through a Jacobian, i.e. when the hybrid frequency is used ($m\geq m_L$), the following adjustment should be applied,
\begin{equation}
		\pi T\sum ... \to \pi T\sum \(1+\frac{k}{2}e^{km-b}\)...
\end{equation}

\begin{figure}
	\begin{center}
		\includegraphics[width=12cm]{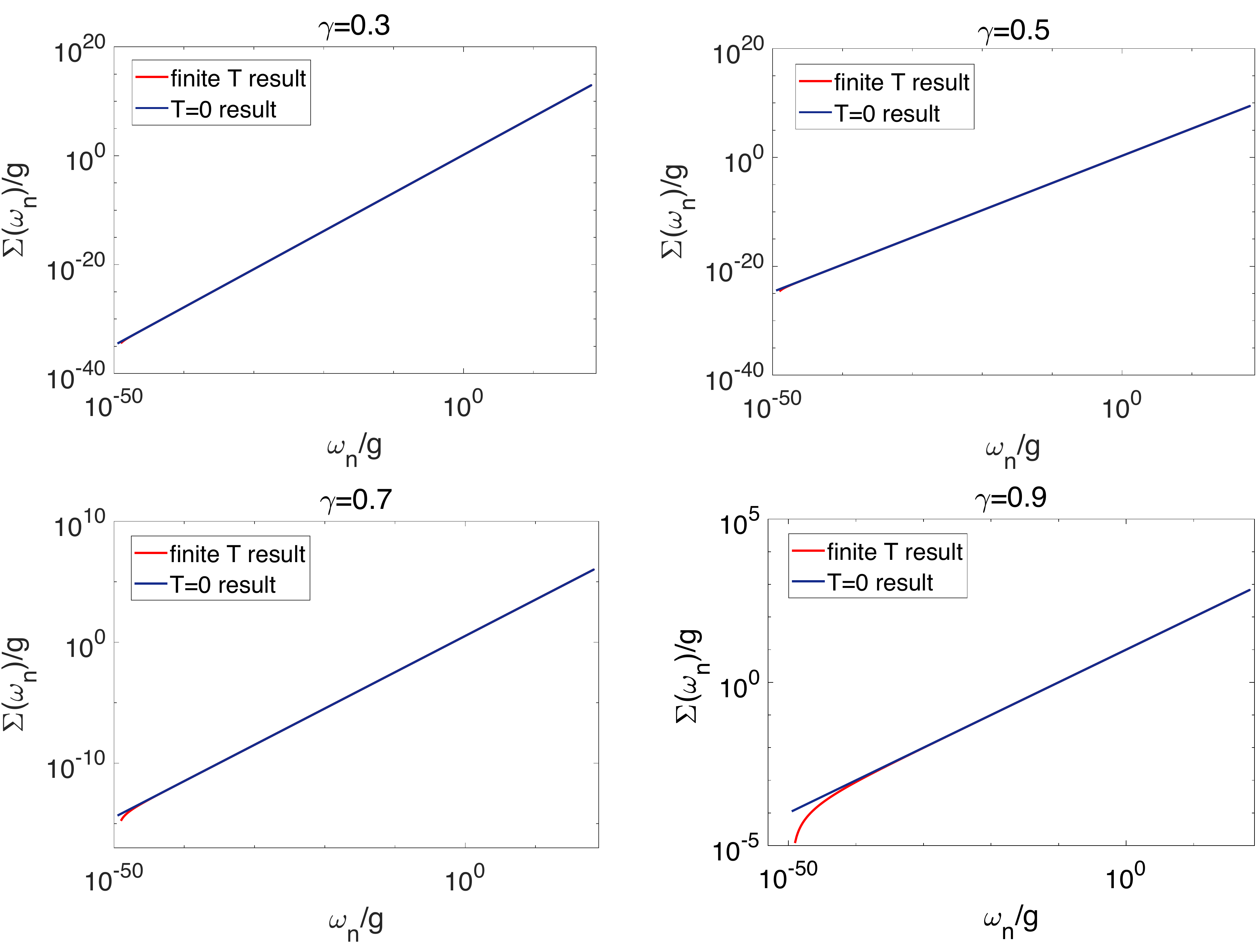}
		\caption{Comparison between self energy obtained though finite temperature($T/g=10^{-50}$) analysis and zero-temperature analysis. In all subplots the red curves represent results obtained via hybrid scaling frequency technique and though the Euler-Maclaurin formula, while the dark blue curves represent $T=0$ analysis $\Sigma(\omega)=\frac{g^\gamma}{1-\gamma}\omega^{1-\gamma}$.}\label{fig:Sigma}
	\end{center}
\end{figure}

On the other hand, the Matsubara summation for the self-energy
is well approximated using the Euler-Maclaurin formula.  In the normal state and at zero bosonic mass $M_b =0$,
  $\Sigma (\omega_m) =  {\bar g}^\gamma (2\pi T)^{1-\gamma} H_{m, \gamma}$, Eq. (\ref{ss_111_a}), where
$H_{m,\gamma}$ is a generalized harmonic number.  Although $H_{m,\gamma}$ is a tabulated  function in some computation libraries, we note that it is well approximated by
\begin{equation}
		\begin{aligned}
			&H_{m,\gamma}=\sum_{p=1}^m\frac{1}{p^\gamma}\approx\frac{1}{1-\gamma}\left(m^{1-\gamma} -1\right)+
\frac{1}{2}\left(1+\frac{1}{m^\gamma}\right)\\
				&+\frac{\gamma}{12}\left(1-\frac{1}{m^{\gamma+1}}\right)-\frac{\gamma(\gamma+1)(\gamma+2)}{720}
\left(1-\frac{1}{m^{\gamma+3}}\right)\\
&+\frac{\gamma(\gamma+1)(\gamma+2)(\gamma+3)(\gamma+4)}{30240}\left(1-\frac{1}{m^{\gamma+5}}\right)\\
		\end{aligned}\label{eq:EM}
\end{equation}
 In Fig. \ref{fig:Sigma} we compare our numerical  $\Sigma (\omega_m)$ at
  about the smallest $T$ that we used, to the zero-temperature expression
  $\Sigma(\omega_m)=\frac{g^\gamma}{1-\gamma}\omega_m^{1-\gamma}$. We see that the numerical and the $T=0$ expressions expectedly coincide at Matsubara numbers $m \gg 1$, but differ at $m = O(1)$, and the difference increases as $\gamma$ gets larger.

In the case when the bosonic mass $M_b$  is finite,   the self-energy in the normal state is given by
$\Sigma (\omega_m) =  {\bar g}^\gamma (2\pi T)^{1-\gamma} H_{m, \gamma} (\delta)$, where $\delta = M_b/(2\pi T)$ and
$ {H}_{m,\gamma}({ \delta})=\sum_{p=1}^m\frac{1}{(p^2+\delta^2)^{\gamma/2}}$. This ${H}_{m,\gamma}({ \delta})$ is not a tabulated function, but based on what we know about the case $M_b =0$, we expect  the  Euler-Maclaurin formula to work well. Using this formula, we obtain
\begin{equation}
		\begin{aligned}
			&{H}_{m,\gamma}( \delta)=\sum_{p=1}^m\frac{1}{(p^2+\delta^2)^{\gamma/2}}\\
		&\approx\left(\frac{1}{\delta}\right)^{\gamma}\left(m _2F_1[\frac{1}{2},\frac{\gamma}{2},\frac{3}{2},-\frac{m^2}{\delta^2}]- _2F_1[\frac{1}{2},\frac{\gamma}{2},\frac{3}{2},-\frac{1}{\delta^2}]\right)\\
		&+\frac{1}{2}\left(\frac{1}{(1+\delta^2)^{\gamma/2}}+\frac{1}{(m^2+\delta^2)^{\gamma/2}}\right)\\
		&+\frac{\gamma}{12}\left(\frac{-m}{(m^2+\delta^2)^{\gamma/2+1}}+\frac{1}{(1+\delta^2)^{\gamma/2+1}}\right)\\
		&-\frac{1}{720}\gamma(\gamma+2)\left(\frac{-m^3(\gamma+4)}{(m^2+\delta^2)^{\gamma/2+3}}+\frac{3m}{(m^2+\delta^2)^{\gamma/2+2}}+\frac{\gamma+4}{(1+\delta^2)^{\gamma/2+3}}-\frac{3}{(1+\delta^2)^{\gamma/2+2}}\right)\\
		&+\frac{1}{30240}\gamma(\gamma+2)(\gamma+4)\left(\frac{-m^5(\gamma+6)(\gamma+8)}{(m^2+\delta^2)^{\gamma/2+5}}+\frac{10m^3(\gamma+6)}{(m^2+\delta^2)^{\gamma/2+4}}-\frac{15m}{(m^2+\delta^2)^{\gamma/2+3}}\right.\\
		&\left.+\frac{(\gamma+6)(\gamma+8)}{(1+\delta^2)^{\gamma/2+5}}-\frac{10(\gamma+6)}{(1+\delta^2)^{\gamma/2+4}}+\frac{15}{(1+\delta^2)^{\gamma/2+3}}\right)\\
		\end{aligned}
\end{equation}
We used this form to compute $\Sigma (\omega_m)$ for a finite $M_b$.

All numerical results, reported in this paper, are obtained using Matlab2017a with the default double precision and Wolfram Mathematica 11.1.

\bibliography{finite_T_gamma_less_1}

\end{document}